\UseRawInputEncoding
\documentclass[journal]{IEEEtran}
\usepackage{mathrsfs}
\usepackage[T1]{fontenc}
\usepackage{cite}
\usepackage{psfrag}
\usepackage{subfigure}
\usepackage{url}
\usepackage{stfloats}
\usepackage{array}
\usepackage{amsthm}
\usepackage{booktabs}  
\usepackage{algorithm} 
\usepackage{setspace}
\usepackage{algorithmic} 
\usepackage{amsmath,amsfonts,amsbsy,bm,paralist,theorem,ifthen,color}
\usepackage{graphicx}
\usepackage{placeins}
\usepackage{float}
\usepackage{hyperref}
\usepackage{tabularx,colortbl}
\usepackage{multirow}
\usepackage{multicol}
\usepackage{arydshln}
\usepackage{colortbl}
\usepackage{tipa}
\usepackage{siunitx}
\usepackage{amssymb}

\ifCLASSINFOpdf
\fi
\begin{document}

\title{Artificial Intelligence Enabled Radio Propagation for Communications---Part II: \\Scenario Identification and Channel Modeling}

\author{Chen Huang,~\IEEEmembership{Member,~IEEE,}
        Ruisi~He,~\IEEEmembership{Senior~Member,~IEEE,}
        Bo~Ai,~\IEEEmembership{Senior~Member,~IEEE,}
        Andreas~F.~Molisch,~\IEEEmembership{Fellow,~IEEE,}
        Buon~Kiong~Lau,~\IEEEmembership{Senior~Member,~IEEE,}
        Katsuyuki~Haneda,~\IEEEmembership{Member,~IEEE,}
        Bo~Liu,~\IEEEmembership{Senior~Member,~IEEE,}
        Cheng-Xiang~Wang,~\IEEEmembership{Fellow,~IEEE,}
        Mi~Yang,~\IEEEmembership{Member,~IEEE,}
        Claude~Oestges,~\IEEEmembership{Fellow,~IEEE,}
        Zhangdui~Zhong,~\IEEEmembership{Senior~Member,~IEEE}
               \\ (\emph{Invited Paper})

\thanks{This work is supported by National Key R\&D Program of China under Grant 2020YFB1806903, the National Natural Science Foundation of China under Grant 61922012, U1834210, and 61961130391, the State Key Laboratory of Rail Traffic Control and Safety under Grant RCS2020ZT010, the Fundamental research funds for the central universities under Grant 2020JBZD005 and I20JB0200030, and the China Postdoctoral Science Foundation under Grant 2021M702499. (\emph{Corresponding authors: Ruisi He; Bo Ai.})}
\thanks{C. Huang is with the Purple Mountain Laboratories, Nanjing, 211111, China, and also with the National Mobile Communications Research Laboratory, School of Information Science and Engineering, Southeast University, Nanjing, 210096, China (email: huangchen@pmlabs.com.cn). Part of this work was done when he was a Ph.D candidate at Beijing Jiaotong University.}
\thanks{R. He, M. Yang, and Z. Zhong are with the State Key Laboratory of Rail Traffic Control and Safety, and the Key Laboratory of Railway Industry of Broadband Mobile Information Communications, Beijing Jiaotong University, Beijing 100044, China (email: {ruisi.he, myang, zhdzhong}@bjtu.edu.cn).}
\thanks{B. Ai is with the State Key Lab of Rail Traffic Control and Safety, Beijing Jiaotong University, Beijing 100044, China, with the School of Information Engineering, Zhengzhou University, Zhengzhou 450001, China, and also with Peng Cheng Laboratory, Shenzhen 518055, China (e-mail: aibo@ieee.org).}
\thanks{A. F. Molisch is with the Ming Hsieh Department of Electrical and Computer Engineering, University of Southern California, Los Angeles, CA 90089 (email: molisch@usc.edu).}
\thanks{B. K. Lau is with the Department of Electrical and Information Technology, Lund University, Lund 22100, Sweden (email: bklau@ieee.org).}
\thanks{K. Haneda is with the Department of Radio Science and Engineering, Aalto University, 00076 Aalto, Finland (email: katsuyuki.haneda@aalto.fi).}
\thanks{B. Liu is with the School of Engineering, University of Glasgow, Glasgow G12 8QQ, U.K. (email: Bo.Liu@glasgow.ac.uk).}
\thanks{C.-X. Wang is with the National Mobile Communications Research Laboratory, School of Information Science
and Engineering, Southeast University, Nanjing, 210096, China, and also with the Purple Mountain Laboratories, Nanjing, 211111, China (email: chxwang@seu.edu.cn).}
\thanks{C. Oestges is with the Institute of Information and Communication Technologies, Electronics and Applied Mathematics, Universite Catholique de Louvain, 1348 Louvain-la-Neuve, Belgium (email: claude.oestges@uclouvain.be).}

}


\maketitle

\begin{abstract}
This two-part paper investigates the application of artificial intelligence (AI) and in particular machine learning (ML) to the study of wireless propagation channels. In Part I, we introduced AI and ML as well as provided a comprehensive survey on ML enabled channel characterization and antenna-channel optimization, and in this part (Part II) we review state-of-the-art literature on scenario identification and channel modeling here. In particular, the key ideas of ML for scenario identification and channel modeling/prediction are presented, and the widely used ML methods for propagation scenario identification and channel modeling and prediction are analyzed and compared. Based on the state-of-art, the future challenges of AI/ML-based channel data processing techniques are given as well.
\end{abstract}


\IEEEpeerreviewmaketitle

\section{Introduction}

\IEEEPARstart{I}{n} existing wireless communication systems, the study on propagation channels has always been a fundamental research in system design, performance evaluation, and other related topics \cite{wang20206g}. Simultaneously, machine learning (ML)-based artificial intelligence (AI) techniques have become the key to develop the next generation communication system. AI techniques have been introduced into wireless communications for the last two decades. It has been found in the existing research that the ML-based AI techniques help solve the bottlenecks in conventional methods relating to channel and antenna considerations \cite{he2021wireless, 9395374, 9158524, 8952905, khuwaja2018survey}. In the Part I \cite{PartI}, we presented a comprehensive literature review on ML-based channel characterization and antenna-channel optimization, whereas this paper focuses on ML-based scenario identification and channel modeling/prediction.

Wireless channels are naturally determined by the physical propagation environment and the interaction object within them. Different propagation environments thus lead to different channel characteristics. Consequently, scenario identification is important and beneficial for many applications, e.g., intelligent transportation systems (ITS), localization, and channel modeling. There are two kinds of scenario identification: i) discriminating line-of-sight (LoS) or non-line-of-sight (NLoS) scenarios, i.e., identifying whether there is an LoS propagation path; and ii) identifying the specific scenarios, e.g., urban areas, highway areas, or suburban areas. The former case is important for localization and channel modeling: most of the existing localization methods require an LoS signal, and most channel models are parameterized separately for LoS and NLoS scenarios \cite{8805246, 8080236, 7497564, 6698380, 7098333, 8889517, 1364132, 982448}. For the latter case, it is a fundamental precondition for capacity analysis, physical layer scheduling, etc. Meanwhile, ML methods have been shown to give good performance in data classification, ML-based scenario identification thus becomes a hot topic.

Furthermore, the very nature of channel modeling is to analyze the physical propagation mechanisms and reveal the relationship between the physical environment and the channel properties. As mentioned before, the wireless channels can be characterized by the critical channel parameters, i.e., time delay, received power, and angle\footnote{The MPC's angle usually includes azimuth/elevation of arrival/departure.} of each multipath components (MPCs), and modeled by seeking the correlation between different channel parameters and physical environments, e.g., propagation distance. Then, an accurate channel model can be used to predict the channel properties in other locations, frequencies, or directions, that have not been measured. Inspired by the power of ML, researchers have explored for more than 20 years the possibilities of using ML to characterize/model the channel, e.g., \cite{TAPoverview} introduces some typical ML methods for channel modelings. In this paper, we give a comprehensive investigation of the existing ML-enabled channel modeling and prediction in time, space, and frequency domains, where details can be found in the following sections.

This paper is organized as follows. Section II presents ML-based communication scenario identification, whereas Section III gives a thorough review of ML-based channel modeling and prediction. Section IV provides a discussion of challenges and possible future research avenues for the topics above. Finally, Section V concludes the paper.

\section{Scenario Identification} \label{scenarioID}
This section introduces state-of-the-art research in this field and analyzes the pros and cons of different solutions.

\subsection{LoS/NLoS Identification}
Identification of LoS/NLoS is essential due to its application in localization or channel modeling. However, it is especially difficult for mobile devices, and in particular for dynamic propagation channels which are time-varying and non-stationary. In existing studies, there are mainly three solutions to identify LoS/NLoS scenarios: i) \emph{visual inspection based on video}: LoS/NLoS conditions can be manually and accurately identified based on the recording video, as in \cite{He2016VehicletoVehicleRC}; ii) \emph{designing the measurement campaign to distinguish the scenario inherently}: a more direct way of channel modeling is to individually measure the channels in LoS and NLoS scenarios, e.g., in \cite{Wang2017111}; and iii) \emph{AI-based automatic identification}: LoS and NLoS conditions usually lead to different channel characteristics; they thus can be exploited to identify the LoS/NLoS scenario by using various algorithm, as in \cite{Borrs1998DecisionTF, Rabbachin2006MLTE, Venkatesh2007NonlineofsightII, 8301878, Yu2013NLOSIA,Heidari2008IdentificationOT, 4394450, 4167834,6492826, Zhang2013AnalysisOK, kurtosis, Gven2008NLOSIA, Gaber2014UDPIF, 7218588,8510840, Yang2018ScenarioCO, losnlos2020, Wymeersch2012AML, 6734961, Xiao2015NonLineofSightIA, 7816727, 8968748, 1453516,Nguyen2015MachineLF, AlHajri2018ClassificationOI, 8254052, 7997068, Zheng2020ChannelNI, 6883312, chen2019orbital}. Considering the time-consuming nature of the visual inspection, and the additional efforts needed in designing measurement campaigns to provide LoS/NLoS distinction, AI-based automatic identification becomes the most promising solution.

According to the methodology, AI-based automatic identification can be classified into two categories: \emph{unsupervised identification} and \emph{supervised identification}. The main difference between these two types is, the supervised identification generally requires a training procedure that needs classified data as training data, and the unsupervised identification directly identifies the LoS/NLoS scenario based on the unclassified characteristics. In another word, the supervised identification requires classifying the training data first, whereas the unsupervised identification does not. Both solutions have been drawing a lot of attention.

\subsubsection{Unsupervised Identification}
As early as 1998, \cite{Borrs1998DecisionTF} formulate the NLoS identification problem as a binary hypothesis test where the range measurements are modeled as being corrupted by additive noise, with different probability distributions depending on the hypothesis. Then, a decision-theoretic framework based on the probability density function (PDF) of the time-of-arrival (ToA) was proposed to solve the hypothesis test problem. Similar solutions are also used in \cite{Rabbachin2006MLTE, Venkatesh2007NonlineofsightII, 8301878, Yu2013NLOSIA}.

The solutions above only focus on the received signal strength and the ToA, \cite{Heidari2008IdentificationOT, 4394450, 4167834} further introduce the root mean square (RMS)-delay spread as another key parameter for the LoS/NLoS identification, and perform a likelihood function test to select the most probable hypothesis. The defined likelihood functions are the simplified Bayesian alternative to the conventional hypothesis testing. In addition, the overlapping area\footnote{the MPCs that have the propagation distance approximating to the LoS distance} between the propagation distance (delay) and the LoS distance is adopted as a testing principle for LoS/NLoS identification in \cite{6492826}, whereas the kurtosis is used for the identification in \cite{Zhang2013AnalysisOK, kurtosis, Gven2008NLOSIA}. Similarly, \cite{Gaber2014UDPIF} studies the automatic LoS/NLoS identification for indoor positioning, where a hybrid time-power test is defined to perform a binary likelihood-ratio test, \cite{7218588} defines a particular feature, named PhaseU, based on the phase information of each subcarrier from channel impulse response (CIR) to identify the LoS/NLoS conditions, whereas the relationship between the Ricean-K-factor and the measured distance is investigated in \cite{8510840} to distinguish LoS and NLoS conditions. Besides hypothesis testing, clustering algorithms have been also studied for the identification, e.g., \cite{Yang2018ScenarioCO} applies the K-Means algorithm to analyze the multi-dimension attributes of the wireless channels for the scenario identification.

\begin{figure}[!t]
    \centering
    \includegraphics[width=0.49 \textwidth]{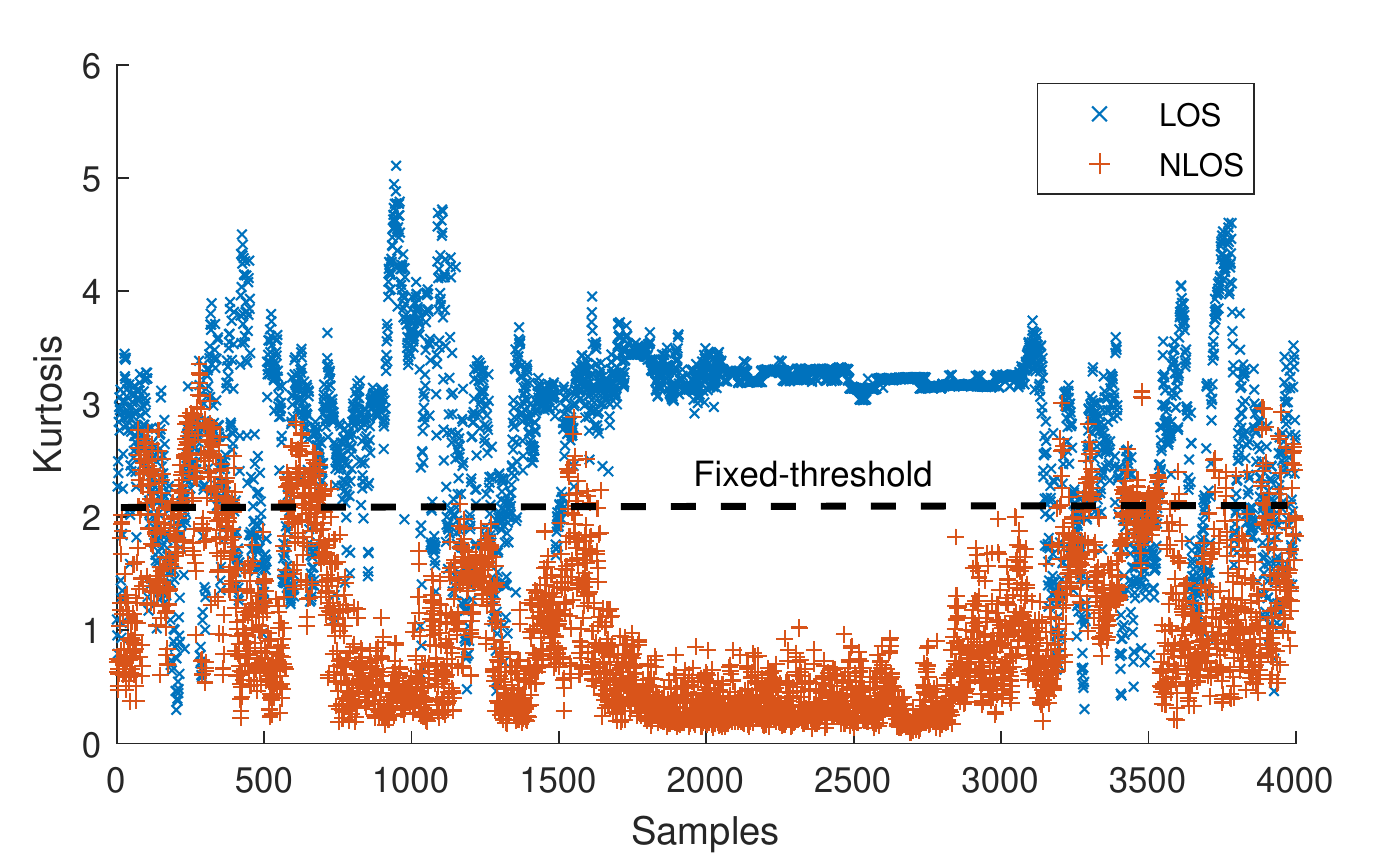}
    \caption{Fix-threshold of kurtosis to dividing the LoS and NLoS channel data in \cite{8968748}.}
    \label{fig_kurtosis}
    \end{figure}

The existing unsupervised identification generally focuses on analyzing the relationship and difference between several statistical channel characteristics, e.g., ToA, RMS-delay spread, and received signal strength, but all use a fixed threshold (obtained by hypothesis testing)\footnote{The hypothesis testing-based methods are usually considered as statistical methods rather than AI methods in some research.} to separate the data to LoS/NLoS scenario. Due to the distinct difference between the LoS and NLoS propagation process, the fixed threshold works well for static scenarios. However, once it comes to time-varying/dynamic scenarios, different thresholds are required for different environments, it thus is hard to find the boundary of the key characteristics between LoS and NLoS channels. In other words, it is difficult to use a fixed threshold to perfectly separate the LoS and NLoS channel data. As shown in Fig. \ref{fig_kurtosis}, the green and blue marks represent the kurtosis of the LoS and NLoS data collected from the channel measurement campaign \cite{Wang2017HighResolutionPE}. It can be found that the LoS and NLoS data overlap each other for some measurement samples, therefore, the fixed threshold generated by the hypothesis testing cannot accurately distinguish the LoS/NLoS scenarios. In this case, supervised-learning methods become an alternative solution for the LoS/NLoS identification.

\subsubsection{Supervised Identification}
Unsupervised identification performs well only for linear classification problems, whereas the time-varying channel data usually show non-linear distribution. On the other hand, supervised identification can learn from the key characteristics and distinguish two non-linear distributions. Therefore, methods based on Support Vector Machine (SVM), Relevance Vector Machine (RVM), Decision Tree, Random Forest, and Artificial Neural Network (ANN) have been widely used for the data classification problem.

SVMs have been proved to have a great ability for non-linear classification problems, especially for binary classification problems. They thus have been widely exploited for the LoS/NLoS identification problem, e.g., \cite{losnlos2020, Wymeersch2012AML, 6734961, Xiao2015NonLineofSightIA, 7816727, 8968748}. The key idea of the SVM is to use a proper kernel function, e.g., Gaussian radial basis function or Polynomial Kernel, and project the data into a higher dimension, where a hyperplane that approximates the metric of interest can be found by training to separate the data. Specifically, the choice of the feeding feature vector for SVM is critical for the training performance. The least-squares SVM (LS-SVM) \cite{Suykens2004LeastSS} is exploited in \cite{losnlos2020} which is trained by the channel features extracted from the channel data, i.e., the received signal strength, maximum amplitude, rising time, mean excess delay, RMS-delay spread, and kurtosis. The LS-SVM uses an especially simple optimization to learn the weights in the SVM to avoid the quadratic programming problem. Based on the channel features used in \cite{losnlos2020}, the entropy and variance of the CIR in the training feature are further exploited in \cite{6734961} to improve the classification performance of the SVM. Instead of using conventional channel characteristics, \cite{Xiao2015NonLineofSightIA} trains the LS-SVM by using statistical features of the data, i.e., mean and standard deviation, skewness, and kurtosis of power delay profile (PDP), Ricean-K-factor, goodness of fit (with a Rician distribution), and the distributions of all the features above. Several typical clustering algorithms, e.g., K-nearest neighbor (KNN), K-means, and self-organizing maps, are compared with SVM in \cite{7816727}, where the SVM generally achieves the highest LoS/NLoS identification accuracy. Similarly, the SVM is also adopted in \cite{lyu2019target} to identify LoS/NLoS condition based on power-angle-spectrum (PAS) features.

Besides the SVM, the RVM is also widely used for classification problems.  Different from the SVM, the RVM is developed based on Bayesian learning forms \cite{1453516}, which requires fewer training samples (compared to the SVM) but also leads to longer training time (with the same amount of training data). An RVM-based NLoS identification is proposed in \cite{Nguyen2015MachineLF}, where channel features similar to those used in \cite{losnlos2020} are adopted.

The Decision Tree, which is constructed based on the training data, is another well-known method for binary classification problems. Therefore, the KNN, SVM, and Decision Tree are compared in \cite{AlHajri2018ClassificationOI}. Trained by the channel transfer function and frequency coherence function, the KNN achieves the best accuracy, whereas the SVM and the Decision Tree show similar performance. Considering that a single Decision Tree may cause an overfitting problem, a random forest consists of multiple Decision Trees trained by different features. Ref. \cite{8254052} trains the Random Forest with the channel features used in \cite{losnlos2020}, and achieves fairly good performance.

Recently, ANNs have also shown great potential for classification problems. A backpropagation (BP) ANN method is adopted in \cite{7997068} and trained by similar statistic features used in \cite{Xiao2015NonLineofSightIA} to identify the LoS/NLoS scenario. Considering the good performance of convolutional neural networks (CNN) for image processing, a CNN-based LoS/NLoS identification algorithm is proposed in \cite{Zheng2020ChannelNI}, where the network is trained directly using the PAS.

\begin{table*}[]
\caption{Summary of the Use of Channel Features for LoS/NLoS Identification.}\label{tb_channelfeature}
\center
\begin{tabular}{|l|l|c|c|m{150pt}|}
\hline
\multirow{2}{*}{Characteristic} & \multirow{2}{*}{Channel Features} & \multicolumn{2}{c|}{Characteristic Type} & \multirow{2}{*}{Existing Works} \\ \cline{3-4}
                                &                                   & Static    & Time-Varying  &                                 \\ \hline
\multirow{5}{*}{Amplitude-based}  &$\bullet$ Maximum Amplitude                 &         $\surd$         &                       & [19], [35]--[37], [40], [42], [52]  \\ \cdashline{2-5}[0.8pt/2pt]
                                & $\bullet$ Received Amplitude                    &             $\surd$     &                & [21], [22], [24], [25], [29], [31], [35]--[38], [42]--[44], [47]     \\ \cdashline{2-5}[0.8pt/2pt]
                                & $\bullet$ Ricean-K-factor                   &        $\surd$          &                       &  [33], [40], [44], [45], [52]\\ \cdashline{2-5}[0.8pt/2pt]
                                & $\bullet$ Kurtosis                          &          $\surd$        &                       &  [28]--[30], [35]--[38], [40], [42], [45], [52]\\ \cdashline{2-5}[0.8pt/2pt]
                                & $\bullet$ Skewness                          &       $\surd$           &                       &  [38], [40], [45], [52] \\ \hline
\multirow{5}{*}{Delay-based}    & $\bullet$ Propagation Distance              &        $\surd$          &                       &  [19], [27], [52]     \\ \cdashline{2-5}[0.8pt/2pt]
                                & $\bullet$ Mean Excess Delay                 &        $\surd$          &                       & [30], [31], [35]--[37], [42], [44], [52] \\ \cdashline{2-5}[0.8pt/2pt]
                                & $\bullet$ Rising Time                       &        $\surd$          &                       &  [35]--[37], [40], [42], [52]  \\ \cdashline{2-5}[0.8pt/2pt]
                                & $\bullet$ RMS-Delay Spread                  &          $\surd$        &                       & [21], [24]--[26], [29], [30], [35]--[37], [40], [42], [52]  \\ \cdashline{2-5}[0.8pt/2pt]
                                & $\bullet$ ToA                               &      $\surd$            &                       &   [19]--[26]    \\ \hline
Power-angular-based             & $\bullet$ PAS                               &        $\surd$          &                       &       [40], [46] \\ \hline
Power-phase-based               & $\bullet$ PhaseU                            &          $\surd$        &                       &   [32]             \\ \hline
\multirow{2}{*}{Angular-based}  &$\bullet$ Angular Difference                &                  &         $\surd$              &      [40], [51], [52]          \\ \cdashline{2-5}[0.8pt/2pt]
                                & $\bullet$ Angular Variant                   &                  &             $\surd$          &       [40]      \\ \hline
\end{tabular}
\end{table*}

Overall, the SVM, Random Forest, and ANN show the most potential for accurate LoS/NLoS identification. Therefore, these three methods are compared in \cite{8968748} trained by the typical channel features, as used in \cite{losnlos2020}, the time-varying channel features, i.e., angular difference and angular variant defined in \cite{8968748}, and the PAS directly. From the comparison evaluation, the Random Forest method outperforms the others in most cases, but the ANN achieves the best performance on PAS-training samples. Furthermore, it also shows that the time variations of the channel features are crucial for the LoS/NLoS identification. Similarly, the angle information in addition to the channel features adopted in \cite{losnlos2020} is applied in \cite{huang2019angular}, which significantly reduces the identification error. Furthermore, the selection of different kernel functions of SVM for LoS/NLoS identification is investigated in \cite{huang2019research}, where Gaussian kernel-based SVM achieves the highest identification accuracy.

In addition, Hidden Markov Model (HMM) introduced in Part I-Section II-B \cite{PartI} has also been adopted to learn the PDP distribution patterns. Ref. \cite{6883312} assumes the Tx and Rx are discrete grid points that can be further divided into distinct non-overlapped cells with different channel conditions; these are then used to train an HMM classifier to identify the LoS/NLoS scenario.

Compared with the unsupervised solution, the supervised LoS/NLoS identification improves the identification accuracy significantly. The main drawback is that it requires pre-classification for the training procedure, which may need a human inspection to identify the training labels first or a specific measurement campaign to collect the LoS and NLoS channel data individually. Besides, as shown by the existing research, the selection of the channel features for training is critical to the final identification performance. Table. \ref{tb_channelfeature} summarizes the widely used channel characteristics for the training features. It can be found that the received amplitude, kurtosis of PDP, and RMS-delay spread are most widely used in the existing studies, since the LoS and NLoS conditions have a major influence on the power and delay attributes, whereas the maximum amplitude, ToA, Ricean-K-factor, and mean excess delay are also frequently adopted.

\subsection{Communication Scenario Identification}

As an extension of the LoS/NLoS identification, communication scenario identification is another similar but separate research topic. The wireless communication's service quality is easily affected by the change of the surrounding environment, especially for dynamic devices, such as vehicular communications or railway communications. These influences are reflected in the channel model, physical layer algorithm, and network layer design. In terms of channel model, \cite{fodor2019supporting} points out that there are obvious differences in path loss between different scenes, for example, the path loss exponents of highways are larger than those of urban areas. The physical layer algorithms also need intelligent scenario identification to adjust algorithm parameters or configurations in real-time according to the environment. In addition, real-time scenario identification contributes to the adaptive adjustment of network architecture. For integrated networks with multiple networking modes, scenario identification is helpful for devices to predict the changes of network state and make necessary adjustments. To sum up, if channel scenarios can be identified accurately and in real-time, the appropriate channel model and transmission mode can be selected to adapt to the current propagation environment, which is helpful to improve the performance and reliability of wireless communication systems.

An identification method is proposed in \cite{6827847} for high-speed railway communication systems based on geographic information systems (GIS). However, these GIS-based methods are not suitable for indoor scenarios or dynamic environments because of the difficulties of localization due to unexpected situations such as special scenarios or sudden obstructions. The physical environment of different scenarios leads to different channel features, so channel features can be used as the basis for scenario identification. The advantage of this is that the actual communication process often needs to continuously monitor the channel state anyway, so there is no obvious additional required system complexity and communication overhead. Moreover, channel features come from the physical environment, reflecting the essential attributes of scenarios, which are not easily affected by light, temperature, weather, and other factors. The key of channel-feature-based scenario identification is to establish the intrinsic mapping relationship between scenarios and corresponding channel features. However, the correspondence between channel features and scenarios is complex. Generally, it is impossible to distinguish all scenarios accurately by a single channel feature. In other words, scenario changes will affect a variety of channel features at the same time, which makes it difficult to directly describe the mapping relationship between channel features and scenarios. Therefore, a feasible solution is to model the nonlinear relationship between channel features and scenario categories through appropriate ML methods. Ref. \cite{alhajri2018classification} suggests a classification method of the indoor environment, which is based on real-time measurement data in a real environment and realizes the classification and identification of different indoor environments, e.g., lab, lobby, and office, by using the weighted KNN method. An ANN-based scenario identification model is presented in \cite{yang2020machine}.  Furthermore, the model configuration scheme, including channel feature type, training data size, and neural network structure, is explored and presented which can make the proposed identification model achieve optimal performance. The validation results show that the identification accuracies are above 98\% in some typical scenarios.

Apparently, there are many similarities between the LoS/NLoS identification and the communication scenario identification: they have the similar mathematical problem, which is a classification task based on the input information, i.e., the fingerprint information (as listed in Table \ref{tb_channelfeature}) of the existing propagation channels. In this sense, developing ML-based LoS/NLoS identification algorithms and ML-based communication scenario identification algorithms can share similar strategies, e.g., SVM or Random forest. On the other hand, from the application point of view, these two techniques have different application environments. The LoS/NLoS identification is mainly conducted for terminal localization/positioning, whereas the communication scenario identification is usually conducted for system-level resource allocation or channel modeling. Therefore, these two techniques share a similar developing principle but have different implementations. Additionally, the LoS/NLoS identification has been studied for several years, whereas the communication scenario identification has recently received much attention. From this point of view, it is possible that these two sub-topics maybe considered jointly in future, as a powerful tool of environment awareness for the Integrated Sensing and Communication (ISAC).

Based on the existing research, we can summarize the two critical aspects for developing the ML-based scenario identification method: a well-selected and appropriate set of features, as the identification basis; and an efficient and accurate identification algorithm, to identify the scenario based on the given features. As proved in \cite{8968748, huang2019angular, lyu2019target}, the selection of features shows more impacts on the identification accuracy than the selection of identification algorithms. Especially considering that there are many kinds of statistical properties of wireless channels, e.g., \emph{the first-order statistics} like mean, PDF, and CDF of received power, and \emph{the second-order statistics} like delay/angular spreads and correlation functions, and only a few of them contribute to the identification accuracy whereas the others may bring interference instead. It can be found from the existing experiments that the selection of the features usually depends on the propagation environment, i.e., the delay-related features are more helpful for the identification in indoor communication channels \cite{Heidari2008IdentificationOT, losnlos2020, Wymeersch2012AML}, whereas the angular-related features greatly contribute to the identification in V2V communication channels \cite{huang2019angular, huang2019research}. On the other hand, the selection of identification algorithms not only depends on the type of the features but also depends on the practical experiment. For example, the RVM is usually recommended when the volume of training samples is limited, as in \cite{Nguyen2015MachineLF}; the CNN is usually recommended when the training feature contains a spatial structure, e.g., PAS, as in \cite{Zheng2020ChannelNI}. However, this is only references rather than principles for selecting identification method, the final selection still depends on the experiments based on measurement data. This also leads to another important issue, i.e., the training data collection for the experiments, which is discussed in Section IV-A.

\section{Channel Modeling}\label{SecChannelModeling}

\subsection{Conventional Channel Modeling}

As covered in Section I, channel modeling becomes increasingly more multi-dimensional as the generation of cellular network evolves.

Now, in the 5G and beyond, due to the emergence of, e.g., massive multiple-input multiple-output (massive MIMO), the multi-dimensionality of channels contains more information than single-input single-output (SISO) channels. The added constraints to channel model parameters usually includes i) multiple mobile terminals, which are close to each other; ii) antenna elements of a physically large antenna array at the base station. Those two added constraints are called i) mobile (user) spatial consistency and ii) non-stationarity at the base station, respectively. In the following, the most popularly referenced channel models are introduced, all of which are intended to reproduce channel parameters or parameters that can describe the measured reality of wave propagation.

\subsubsection{Statistical Modeling}
Traditionally, stochastic channel models have been used to describe path loss, shadowing, and small-scale fading, among others. The mathematical formulas defining the models are {\it heuristically} derived from a large number of empirical observations, i.e., {\it massive data of wave propagation}. The formulas have several variables, whose values are in practice limited to physically justifiable ranges; for example, a path loss exponent in free-space can take only a value of $2$ due to the spherical spreading of power radiated from a point source. The existing channel models are therefore a good starting point when developing improved models through computer-assisted model-based learning. Furthermore, providing a physical understanding of parameter values of the improved channel model is always useful to justify the soundness of the model.

These stochastic models, however, suffer from a significant drawback when fulfilling the added constraints for 5G and beyond, e.g., they lack implicit modeling of user spatial consistency and non-stationarity at the base station, defined at the beginning of this section. The introduced pure statistical models are therefore used mainly for link-level analyses where different user links are assumed in the above independent spatial-, time-, and frequency- properties, and moreover base station antennas are arranged so that they all see the same propagation paths.

\subsubsection{Geometrical Modeling}
The above-mentioned drawbacks of purely statistical models have been solved by introducing geometrical information, which relates variations of radio channel responses with locations and movement of radio devices and waves scattering objects in the environment. Geometrical modeling of channels translates the multipath parameters into a set of parameters for a given location of a base station and a mobile device antenna, along with their moving velocity and direction if applicable.

Site-specific propagation modeling is one of the typical geometrical modeling approaches, which focuses on the mathematical characterization of MPCs arising from radio waves interacting with physical objects in the environment, leading to reflection, scattering, diffraction, and penetration. Therefore geometrical modeling of propagation channels inherently allows deriving the channel model parameters with the added constraints defined at the beginning of this section.

An approximate solution of wave propagation can be obtained from ray-tracing (RT), which is based on geometrical optics that uses ray concepts to determine the reflected and refracted fields from a surface. In a different approach, channel modeling based on statistical geometrical information of a specific cellular site has been an interest for planning base station deployment. Seminal works of Ikegami, Walfisch, and Bertoni in the 1980s identify rooftops of office buildings and residential houses as the major wave interacting points in urban macrocellular environments where rooftop base stations serve ground-level outdoor mobile devices. Losses of signal paths pertaining to the wave interaction are estimated by modeling the buildings and houses a series of absorbing screens, e.g.,~\cite{Ikegami84TAP, Walfisch88TAP} and \cite{Bertonibook}. For micro-cellular scenarios where base stations are installed below rooftop level, corners of buildings serve as wave interacting points in street intersections~\cite{COST231}. Therefore a ``typical'' map of an environment can also be created to emulate delay and angular characteristics \cite{molisch2002virtual,kunisch2003ultra, Medbo16CM} of a microcellular scenario. Discrete MPCs are obtained from the map, while scatterers causing weaker paths are statistically distributed.

Different from the site-specific approach, there are also approaches to create scattering environments that do not exist in reality, and are purely determined from a probability distribution of the scatterer locations  \cite{molisch2004generic}. Such imaginary environments however still reflect some reality of measured multipath propagation based on statistics derived from measurements. This type of channel model is called reference or canonical model and has been used to compare different radio communications technologies. Reference models such as the 3GPP TR38.901~\cite{TR38901} and COST2100~\cite{Liu2012TheC2} models and their predecessors (3GPP spatial channel model, WINNER, and COST 259, COST 273, respectively) have different steps to derive the spatial-, time-, delay- and polarization-properties of MPCs.

The main difference between the two model types is the extent to which they rely on coordinates of scattering objects for deriving the multipath parameters. The 3GPP TR38.901 model derives geometrically parameters only for the LoS path, while the angle and delay deviations of the other paths relative to the LoS are determined purely statistically. On the other hand, the COST2100 channel model relies fully on the coordinates of scattering objects to derive multipath parameters. This leads to unique advantages and disadvantages of the two models. The 3GPP model allows straightforward implementation using statistical distributions of large-scale parameters such as delay and angular spreads. However, it requires an extra mechanism to control mobile (user) spatial consistency and non-stationarity at the base station, which is only partly available from proposed additional procedures~\cite{Jaeckel14TAP}.

Spatial consistency is implicitly addressed, on the other hand, in the COST2100 model, e.g., because of the introduced concept of {\it visibility regions} \cite{Molisch2006TheCD}. A visibility region is associated with a single cluster, which is activated for a base-to-mobile link when a mobile device falls into the region~\cite{Flordelis20TWC}.

\subsection{ML-Enabled Channel Modeling and Prediction}

The key idea of the conventional channel modeling is to characterize the MPCs/clusters into mathematical functions, based on stochastic process or deterministic derivation. However, due to the limitations of the channel measurement campaigns, we cannot measure the channel in all kinds of scenarios. Therefore, predicting the channel in unknown environments is essential. In this sense, the good accuracy and flexibility of the ML-based predicting methods inspiring us to explore the ML-based channel modeling/prediction solutions. Furthermore, the ML methods is able to learn and extract the underlying properties from the measured data, which cannot be extracted and described by using conventional modeling method. This is also a critical point to accurately model/predict the channels in complicated environment.

We categorize the existing ML-enabled channel studies into two types: i) \emph{ML-based channel characterization/modeling}, which provides the corresponding channel statistical parameters, e.g., \cite{hornik1989multilayer, gschwendtner1993application, balandier1995170, goodfellow2016deep, kalakh2012neural, zaarour2012comparative, zaarour2013accurate, zaarour2015comparative, hu2020mm, oroza2017machine, yang2019machine, wen2019machine, lv2007novel, sha2008wireless, sun2018channel, ma2008artificial, zhao2019neural, ledig2017photo, yang2019generative}, and ii) \emph{ML-based channel prediction}, which exploits ML methods to learn from the acquired channel information to predict the channel characteristics at \emph{different locations}, \emph{different frequency bands} or \emph{different times}, e.g., \cite{neskovic2002macrocell, romo2006application, zhou2005application, vilovic2009using, azpilicueta2014ray, ferreira2016improvement, huang2018big, ozdemir2014prediction, bai2019prediction, bai2020novel, yang2020deep, zhao2020playback, ratnam2020fadenet, zhang2018air, zhu2019adaptive, jiang2019comparison, liu2006recurrent, routray2011rayleigh, jiang2019recurrent, jiang2018multi, sui2018jointly, potter2008mimo, tong2017long, jiang2020deep, jiang2020recurrent, yuan2019machine, yuan2020machine, mehrabi2019decision, alrabeiah2020deep, ding2014fading, ding2013fading, wang2019ul, arnold2019enabling, yang2020deep2}.

\subsubsection{ML-Based Characterization/Modeling} The key idea is to directly capture the channel characteristics, e.g., path loss or delay spread, by using ML methods. Since the ML algorithm shows good accuracy and efficiency for characterizing the channel, it is thus expected to directly give the model or generate the synthetic channel data.

It has been proved in \cite{hornik1989multilayer} that a three-layer multilayer perceptron (MLP) neural network can approximate an arbitrary continuous multidimensional function to any desired accuracy, which provides a theoretical basis for employing neural networks to approximate channel behaviors as functions of physical/geometrical/bias parameters. As early as 1993, Ref. \cite{gschwendtner1993application} studied the channel characteristics by using the standard Fuzzy Neural Network (FNN) trained by the BP algorithm, where the received signal strength is estimated from the receiving distance. Similarly, the BP-trained neural network is also adopted in \cite{balandier1995170} to estimate the received signal strength for wireless channels. Theoretically, for a BP-based neural network, more hidden layers can lead to better performance. However, a large number of the hidden layers may cause the \emph{vanishing gradient} problem \cite{goodfellow2016deep}, limiting the performance. To avoid the vanishing gradient problem, a single layer-based MLP network is proposed in \cite{kalakh2012neural} to obtain the path loss of ultra wideband (UWB) channels with the frequency band from 0.875 MHz to 10 GHz. Instead of using a generalized MLP network, an RBF-based neural network is exploited in \cite{zaarour2012comparative, zaarour2013accurate} to capture the frequency-dependent path loss of the UWB channels and estimate the received power. The Radial Basis Function (RBF)-based network used in \cite{zaarour2012comparative, zaarour2013accurate} is a specialized case of the MLP network which contains only three layers: an input layer, a hidden layer with a non-linear RBF activation function, and a linear output layer. The RBF-based network and the generalized MLP network are compared in \cite{zaarour2015comparative, hu2020mm} for modeling the path loss of 60 GHz wireless channels, where the RBF-based network achieves better accuracy at the cost of higher computational complexity (because of more neurons in the hidden layers). Besides neural networks, some other ML tools are also adopted and compared for channel characterization. Random Forest, AdaBoost, and KNN, as well as neural networks, are used to estimate the receiving signal strength in \cite{oroza2017machine}, where both the Random Forest and the KNN achieve better performance than the generalized neural network. Random Forest and KNN are further compared in \cite{yang2019machine} for path loss calculation, where Random Forest shows better robustness against noise. This gives us the intuition that other ML methods may fit better than the ``universal'' neural network in some specific cases. Therefore, the MLP network and the SVM network\footnote{The SVM network in \cite{wen2019machine} is a specific neural network with a single hidden layer, where the SVM is used as the active function.}-based path loss calculations are investigated and compared in \cite{wen2019machine}. Specifically, the initial parameters for the SVM network are optimized by applying a Genetic Algorithm (GA). In this case, the GA+SVM network outperforms the widely used MLP network for path loss estimation for 60 GHz channels.

The BP-network introduced above is adopted to simulate Additive White Gaussian Noise (AWGN) channels with Doppler shift in \cite{lv2007novel}. Similarly, the RBF-based network is adopted in \cite{sha2008wireless, sun2018channel} and a generalized MLP network is adopted in \cite{ma2008artificial} to simulate a generalized wireless channel, where the network input and output are the system layout (i.e., frequency points and propagation distance) and the channel impulse response, respectively. A time-varying channel model is given in \cite{zhao2019neural} at 26 GHz for outdoor and indoor environment, which exploits two separate ANNs for modeling the path loss with shadowing and the small scale channel parameters.

From the existing works, the greatest advantage of the ML-based channel model compared to the conventional models is that it supports a more flexible environment which thus leads to higher accuracy for describing the channels than the conventional models. Nevertheless, it has been found in many papers that the accuracy of the trained models is limited by the insufficient training data, i.e., measurement data. The limitation of the training database is also a major problem in the computer science field, where the generative adversarial network (GAN) \cite{ledig2017photo} is proposed to generate the synthetic data (usually generated by random variables and testified by a discriminator) which can well simulate the training data. Following this idea, a GAN-based wireless channel modeling framework is given in \cite{yang2019generative}, where the channel fake samples are generated by latent random variables and compared to channel real samples by using a channel data discriminator, as shown in Fig. \ref{fig_channelModelingPrediction}(b).

\subsubsection{ML-Based Channel Prediction}
With the development of ML algorithms, the accuracy of the ML-based modeling is continually higher, such that, the channel modeling is evolving to channel prediction. The main difference between the channel modeling and the channel prediction is that the former case more focuses on generalized channel characteristics, e.g., the path loss, the shadow fading, or the small scale fading; the latter case more focuses on predicting the specific channel characteristic at a particular condition, e.g., at a specific location (space domain prediction), a specific time (time domain prediction), or a different frequency band (frequency domain prediction). Therefore, the channel prediction is expected to provide channel state information in the future or in different frequency bands. Generally, the channel prediction requires more detailed environment information or channel history information and is able to provide more accurate channel characteristics under the given condition. Hence, the set of input training vectors is critical for ML-based channel prediction.

A coverage estimation method is proposed in \cite{neskovic2002macrocell} for different outdoor scenarios, e.g., urban, suburban, forest, rural, or rivers, where an MLP network is adopted to predict the field strength at different physical positions. In this method, the portion through the terrain (defined in \cite{neskovic2002macrocell}), which contains the geometry map information like the position of Tx/Rx, and physical environment factors, i.e., rolling factor and modified clearance angle for Tx and Rx, are introduced as network input, whereas the output is the received power at Rx side. However, the complicated physical environment brings difficulties to accurately predict the receiving power at a specific position. Compared with the outdoor environment, the indoor environment may be more deterministic and the received power at each position is thus more predictable with detailed scenario information. An indoor coverage prediction method is developed in \cite{romo2006application} based on a three-layer MLP network, where free space pathloss, transmission loss, wave-guiding effect, reflection loss, local reflectors, and shielding effect factors are adopted as network input to estimate the field strength. Similarly, a three-layer MLP network is used in \cite{zhou2005application} for indoor coverage prediction where Tx, Rx, antenna position, antenna gain, max transmit power, average attenuation coefficients, the deviation between the angle of arrival (AoA) and angle of departure (AoD), and physical environment factors (distance between Tx and Rx, number of walls and windows between Tx and Rx, average angle of incidence of walls, visibility factor, frequency point, and variety of people density) are used as network input to predict the received power.

To further improve the coverage estimation accuracy, particle swarm optimization (PSO) is used to improve the MLP network training efficiency and accuracy in \cite{vilovic2009using}, where the position of Tx and Rx and the received power are the network input and output, respectively. On the other hand, the training accuracy is also highly dependent on the training database. However, it is very challenging to build a comprehensive database by actually measuring the channels. Therefore, deterministic channel simulators like ray-launching (RL) and RT \cite{fan2015emulating} can provide detailed and sufficient synthetic channel data for the training process, e.g., an RL-neural network is proposed in \cite{azpilicueta2014ray} for complex indoor coverage estimation, where the Tx and Rx coordinates and the received power are the network input and output, respectively.

The position of the Tx and Rx and the diffraction loss by the Cascade Knife Edge is used in \cite{ferreira2016improvement}  as the input to an ANN for generating the received signal, which is compared with the ITU-R.526 model and shows better accuracy when comparing with measurement data. The position of Tx, Rx, and carrier frequency are adopted in \cite{huang2018big} as the input to output the channel statistical properties, i.e., the received power, RMS delay spread, RMS angle spread. It has been found in \cite{huang2018big} that FNN and RBF-based networks show similar modeling performance, which draws a similar conclusion in Section \ref{scenarioID}-A: Compared with the choice of the specific network model, the quality of the training data set displays a greater impact on the accuracy for the communication modeling/classification problem.

Ref. \cite{ozdemir2014prediction} proposes a propagation loss prediction algorithm by using an ANN, where the Rx location, effective antenna height, and terrain irregularity are used as network input to predict the received power at a different location. An ANN-based satellite communication channel prediction algorithm is proposed in \cite{bai2019prediction} to predict the received power, where the weather information, such as air temperature, humidity, rainfall rate, visibility, relative speed, etc., is incorporated as a physical condition factor for the network. The method in \cite{bai2019prediction} is further developed in \cite{bai2020novel} by replacing the generalized ANN with a Long Short-Term Memory (LSTM) network, and adding a weather condition classification process to improve the prediction accuracy. Similarly, an LSTM-based multimodal deep learning network (MDLN) is proposed in \cite{yang2020deep} for downlink channel prediction, where the received signals and pilots, previous downlink channels, Least-squares (LS) estimation results, user location, current uplink channel, partial downlink channel are exploited as network input.  The Tx and Rx locations and the frequency band are adopted in \cite{zhao2020playback} as input to train an ANN network to predict the amplitude, delay, phase, and cross polarization ratio at each location point, and thus can ``playback'' the MIMO channels. A similar approach was taken in \cite{levie2021radiounet}. Instead of using LSTM, a CNN-based pathloss and shadowing prediction is proposed in \cite{ratnam2020fadenet}, where the coordinates of Rx and Tx, the physical environment information, i.e., terrain height, building height, and foliage height, and the visibility condition (LoS/NLoS) are used as input to predict the received power at each position. Instead of using the generalized/specialized ANN, \cite{zhang2018air} exploits the Random Forest and KNN method to build a prediction model for unmanned aerial vehicle channels, which requires fewer initial parameters, i.e., propagation distance, altitude of Tx and Rx, visibility condition (LoS/NLoS), and link elevation angle.

     \begin{figure*}[!t]
    \centering
    \includegraphics[width=0.85 \textwidth]{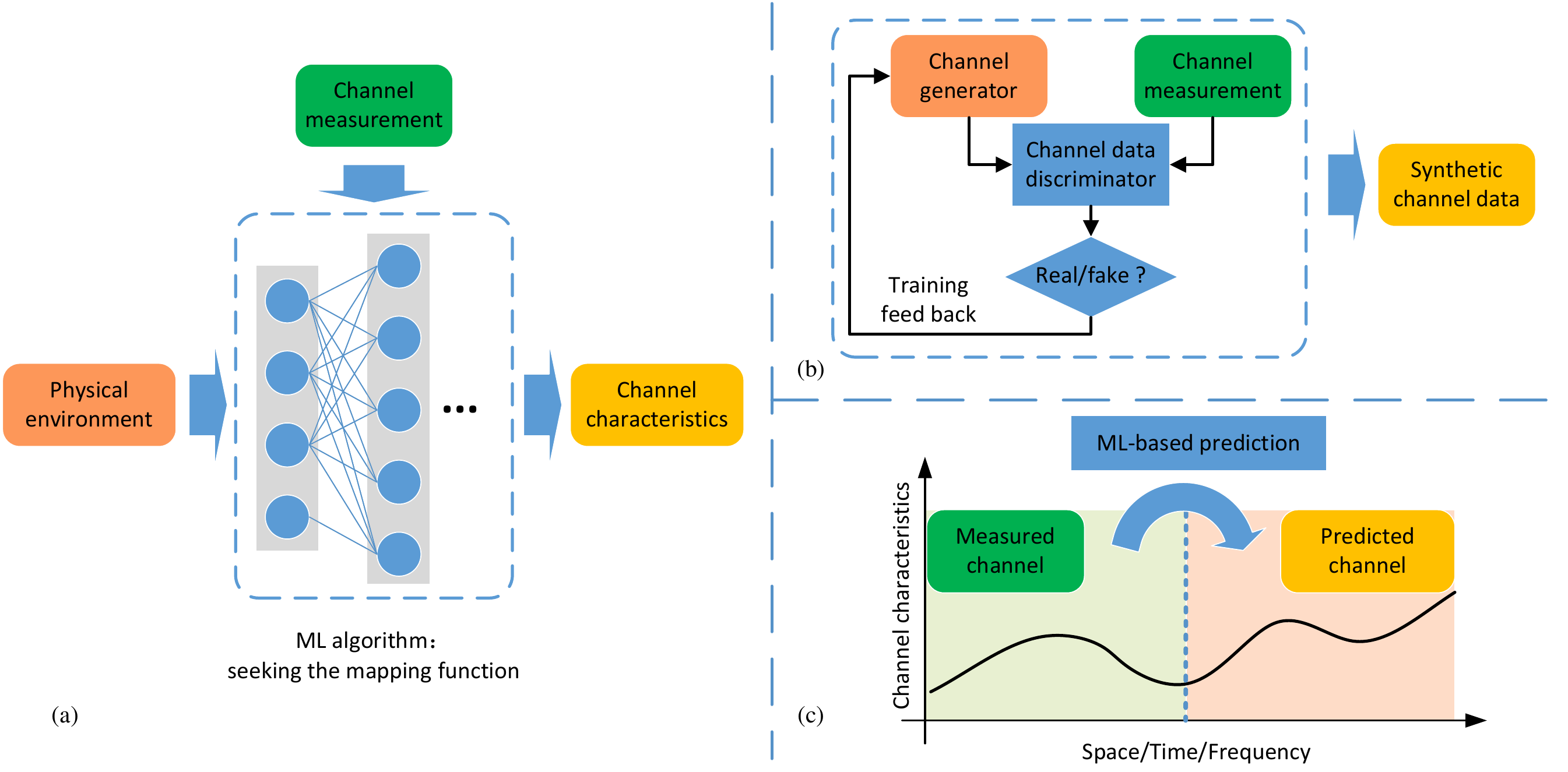}
     \caption{Key ideas of the ML-based (a) channel modeling, (b) channel simulation, and (c) channel prediction.}
    \label{fig_channelModelingPrediction}
    \end{figure*}

On the other hand, many existing channel estimation studies more focus on predicting the channel at a different time for the Time-Division Duplexing (TDD) system or different frequency band for the Frequency Division Duplexing (FDD) system, rather than different locations\footnote{In some cases, locations and times are highly related to each other in time-varying scenarios.}. For most cases, the channel matrixes $\mathbf{H}$ in history (the history time or the known frequency band) are often used as the network input, whereas the network output is the channel matrix $\mathbf{H}$ in the present time or for the unknown frequency band. Following this idea, a recurrent neural network (RNN)-based channel prediction is proposed in \cite{zhu2019adaptive} to predict the channel state information at different times for the time-varying channels. The RNN and Kalman filter are compared in \cite{jiang2019comparison} for predicting the channel matrixes $\mathbf{H}$ in future time, whereas the extended Kalman filter and the decoupled extended Kalman filter are used in \cite{liu2006recurrent} to train the RNN for channel matrix prediction for SISO system. A similar method is also used in \cite{routray2011rayleigh} for MIMO flat-fading channels prediction. A frequency-domain RNN predictor that deals with a frequency-selective MIMO channel as a set of parallel flat-fading sub-carriers is proposed in \cite{jiang2019recurrent}. Ref. \cite{jiang2018multi} proposes real-valued RNNs to implement multi-step predictors for long-term prediction, and further verifies its effectiveness in a TAS system to lower complexity. The PSO method in \cite{vilovic2009using} is also adopted in \cite{potter2008mimo} to help train the RNN network for channel prediction. A jointly optimized extreme learning machine (JO-ELM) for short-term prediction is delivered in \cite{sui2018jointly}, whereas the long-term prediction method based on LSTM is presented in \cite{tong2017long, jiang2020deep, jiang2020recurrent}. In contrast to simply using channel matrix \textbf{H} in history as input, Refs. \cite{yuan2019machine, yuan2020machine} argue for applying CNN to extract channel state information (CSI) and present a CNN-RNN architecture for CSI aging. A decision-directed estimation with deep-FNN based channel prediction is built in \cite{mehrabi2019decision} for MIMO transmission, whereas the ANN-based deep learning is also used in \cite{alrabeiah2020deep} to estimate the beamforming vector instead of the channel matrix \textbf{H} with the input of user position and the channel matrix \textbf{H} history. Instead of using the RNN, \cite{ding2014fading, ding2013fading} deliver a CSI predictive method by means of combing a multi-layer complex-valued neural network (CVNN) with the chirp Z-transform. For the FDD system, \cite{wang2019ul} delivers a hybrid of a CNN and LSTM to extract the downlink CSI according to that of uplink channels assuming strong channel correlation, whereas a deep learning-based extrapolation approach is proposed in \cite{arnold2019enabling} which infers the downlink CSI by solely observing uplink CSI on an adjacent frequency band. A deep transfer-based meta-learning (DT-meta-learning) algorithm is proposed in \cite{yang2020deep2} for downlink channel estimation which trains the network by alternating inner-task and across-task updates and then adapts to a new environment with a small number of labeled data.

Overall, the ML-based channel characterization/modeling and prediction mostly relies on different strong points of the ML methods: the former case generally uses the ML methods for seeking the mapping function between the environment and the channel properties or directly emulate the measurement data, as shown in Fig. \ref{fig_channelModelingPrediction}(a) and (b), respectively, whereas the latter case usually uses the ML method to learn the history of channels\footnote{The history here is not only in time dimension but also includes frequency dimension.} and predict/extrapolate the future/unknown channels, as shown in Fig. \ref{fig_channelModelingPrediction}(c). It can be found in the state-of-the-art studies, most of the ML-based channel prediction algorithms are proposed based on the ANN, which have offline training process and low computational complexity online implementations, but with different network structures. This leads to the major challenges of developing channel prediction methods: i) what type of ANN should be used? and ii) what input and output should be used for training? Facing these two challenges, Table \ref{tb_channelprediction} summarizes the typical ML-based channel prediction methods. It is noteworthy that the material factors, geometry map, and propagation factors may refer to different particular parameters in each work, and the detail of each factor can be found in each reference.

It has been found from the existing research that embodying some prior knowledge into neural network architectures usually induces good generalization \cite{lu2021learning}. This inductive bias has been reflected in many networks, such as CNN for processing the data that have a certain spatial structure, e.g., PAS; and RNN for the data that have a sequential structure, e.g., PDP. This is a general guideline to designing/developing any ML-based applications for the communication system, the network model should be carefully selected according to the characterization of the processed data, e.g., the CNN-based methods are recommended to extract the PAS/channel matrix feature in each snapshot; whereas the RNN-based methods are recommended to process a period time of channel data. This is a direction for selecting a proper ML method, in particular an ANN method. There are still massive kinds of CNNs (e.g., LeNet, AlexNet, VGG-Net, GoogLeNet, and ResNet) and RNNs (e.g., LSTM, Gate Recurrent Unit, and their variants), which needs further experiment to evaluate each method.

\begin{table*}[]
\caption{Summary of the ML-based channel prediction.}\label{tb_channelprediction}
\center
\begin{tabular}{|l|l|l|l|l|}
\hline
Domain                                                                                                      & Existing works           & Machine learning tools                                                           & Method input                                                                                                      & Method output                                                                       \\ \hline
\multirow{14}{*}{\begin{tabular}[c]{@{}c@{}}Space \\ domain \\ prediction\end{tabular}}                     & \cite{neskovic2002macrocell}& \multirow{12}{*}{MLP}                                                       & \begin{tabular}[c]{@{}l@{}}$\bullet$ Material factors\\ $\bullet$ Geometry map\end{tabular}                                          & $\bullet$ Received power                                                                      \\ \cline{2-2} \cdashline{4-5}[1.2pt/1pt]
                                                                                                            & \cite{romo2006application}     &                                                                                 & \begin{tabular}[c]{@{}l@{}}$\bullet$ Propagation factors\\ $\bullet$ Geometry map\end{tabular}                                      & $\bullet$ Received   power                                                                    \\ \cline{2-2} \cdashline{4-5}[1.2pt/1pt]
                                                                                                            & \cite{zhou2005application}     &                                                                                 & \begin{tabular}[c]{@{}l@{}}$\bullet$ Geometry map\\ $\bullet$ Transmit power\\ $\bullet$ Propagation factors\\ $\bullet$ Frequency band;\end{tabular} &$\bullet$ Received   power                                                                    \\ \cline{2-2}\cdashline{4-5}[1.2pt/1pt]
                                                                                                            & \cite{vilovic2009using}        &                                                                                 & $\bullet$ Tx and Rx position                                                                                                & $\bullet$ Received   power                                                                    \\ \cline{2-2} \cdashline{4-5}[1.2pt/1pt]
                                                                                                            & \cite{azpilicueta2014ray}      &                                                                                 & $\bullet$ Tx and Rx position                                                                                                & $\bullet$ Received   power                                                                    \\ \cline{2-2} \cdashline{4-5}[1.2pt/1pt]
                                                                                                            & \cite{ferreira2016improvement} &                                                                                 & \begin{tabular}[c]{@{}l@{}}$\bullet$ Distance between Tx and Rx\\ $\bullet$ Propagation factors\end{tabular}               & $\bullet$  Received power                                                                    \\ \cline{2-2} \cdashline{4-5}[1.2pt/1pt]
                                                                                                            & \cite{huang2018big}            &                                                                                 & \begin{tabular}[c]{@{}l@{}}$\bullet$  Geometry map\\ $\bullet$ Frequency band\end{tabular}                                                & \begin{tabular}[c]{@{}l@{}}$\bullet$  Received power\\ $\bullet$  Delay/angular spread\end{tabular}       \\ \cline{2-2} \cdashline{4-5}[1.2pt/1pt]
                                                                                                            & \cite{ozdemir2014prediction}   &                                                                                 & \begin{tabular}[c]{@{}l@{}}$\bullet$  Geometry map\\ $\bullet$  Material factors\end{tabular}                              & $\bullet$  Received power                                                                      \\ \cline{2-2} \cdashline{4-5}[1.2pt/1pt]
                                                                                                            & \cite{bai2019prediction}       &                                                                                 & \begin{tabular}[c]{@{}l@{}}$\bullet$  Weather factors\\ $\bullet$  Geometry map\\ $\bullet$  Propagation factors\end{tabular}                   & $\bullet$ Channel excess attenuation                                                          \\ \cline{2-2}\cdashline{4-5}[1.2pt/1pt]
                                                                                                            & \cite{bai2020novel}            &                                                                                 & \begin{tabular}[c]{@{}l@{}}$\bullet$   Weather factors\\ $\bullet$  Geometry map\\ $\bullet$  Propagation factors\end{tabular}                   & $\bullet$  Channel excess attenuation                                                          \\ \cline{2-2}\cdashline{4-5}[1.2pt/1pt]
                                                                                                            & \cite{zhang2018air}            &                                                                                 & \begin{tabular}[c]{@{}l@{}}$\bullet$ Geometry map\\ $\bullet$ Material factors\end{tabular}                                         &$\bullet$  Received power                                                                      \\ \cline{2-2} \cdashline{4-5}[1.2pt/1pt]
                                                                                                            & \cite{zhao2020playback}        &                                                                                 & \begin{tabular}[c]{@{}l@{}}$\bullet$  Geometry map\\ $\bullet$  Frequency band\end{tabular}                                           & \begin{tabular}[c]{@{}l@{}}$\bullet$  Amplitude, delay, phase\\ $\bullet$  Cross polarization\end{tabular} \\ \cline{2-3}\cdashline{4-5}[1.2pt/1pt]
                                                                                                            & \cite{yang2020deep}            & LSTM-based MDLN                                                                 & \begin{tabular}[c]{@{}l@{}}$\bullet$ H matrix in history\\$\bullet$  LS estimation results\\ $\bullet$ Geometry map\end{tabular}             & $\bullet$  Downlink CSI                                                                    \\ \cline{2-3}\cdashline{4-5}[1.2pt/1pt]
                                                                                                            & \cite{ratnam2020fadenet}       & CNN                                                                             & \begin{tabular}[c]{@{}l@{}}$\bullet$ Geometry map\\ $\bullet$ Material factors\end{tabular}                                         & \begin{tabular}[c]{@{}l@{}}$\bullet$  Large scale fading\\  in each position\end{tabular}      \\ \hline
\multicolumn{1}{|c|}{\multirow{17}{*}{\begin{tabular}[c]{@{}c@{}}Time \\ domain\\ prediction\end{tabular}}} & \cite{zhu2019adaptive}         & \multirow{6}{*}{RNN}                                                            & \multirow{16}{*}{$\bullet$ CSI in history}                                                                             & \multirow{16}{*}{$\bullet$ CSI for future}                                               \\ \cline{2-2}
\multicolumn{1}{|c|}{}                                                                                      & \cite{liu2006recurrent}        &                                                                                 &                                                                                                                   &                                                                                     \\ \cline{2-2}
\multicolumn{1}{|c|}{}                                                                                      & \cite{routray2011rayleigh}     &                                                                                 &                                                                                                                   &                                                                                     \\ \cline{2-2}
\multicolumn{1}{|c|}{}                                                                                      & \cite{jiang2019recurrent}      &                                                                                 &                                                                                                                   &                                                                                     \\ \cline{2-2}
\multicolumn{1}{|c|}{}                                                                                      & \cite{potter2008mimo}          &                                                                                 &                                                                                                                   &                                                                                     \\ \cline{2-2}
\multicolumn{1}{|c|}{}                                                                                      & \cite{jiang2018multi}          &                                                                                 &                                                                                                                   &                                                                                     \\ \cline{2-3}
\multicolumn{1}{|c|}{}                                                                                      & \cite{ding2014fading}          & \multirow{2}{*}{CVNN}                                                           &                                                                                                                   &                                                                                     \\ \cline{2-2}
\multicolumn{1}{|c|}{}                                                                                      & \cite{ding2013fading}          &                                                                                 &                                                                                                                   &                                                                                     \\ \cline{2-3}
\multicolumn{1}{|c|}{}                                                                                      & \cite{jiang2019comparison}     & Kalman filter/RNN                                                               &                                                                                                                   &                                                                                     \\ \cline{2-3}
\multicolumn{1}{|c|}{}                                                                                      & \cite{sui2018jointly}          & JO-ELM                                                                           &                                                                                                                   &                                                                                     \\ \cline{2-3}
\multicolumn{1}{|c|}{}                                                                                      & \cite{tong2017long}            & \multirow{3}{*}{RNN(LSTM)}                                                           &                                                                                                                   &                                                                                     \\ \cline{2-2}
\multicolumn{1}{|c|}{}                                                                                      & \cite{jiang2020recurrent}      &                                                                                 &                                                                                                                   &                                                                                     \\ \cline{2-2}
\multicolumn{1}{|c|}{}                                                                                      & \cite{jiang2020deep}           &                                                                                 &                                                                                                                   &                                                                                     \\ \cline{2-3}
\multicolumn{1}{|c|}{}                                                                                      & \cite{yuan2019machine}         & \multirow{2}{*}{CNN+RNN}                                                        &                                                                                                                   &                                                                                     \\ \cline{2-2}
\multicolumn{1}{|c|}{}                                                                                      & \cite{yuan2020machine}         &                                                                                 &                                                                                                                   &                                                                                     \\ \cline{2-3}
\multicolumn{1}{|c|}{}                                                                                      & \cite{mehrabi2019decision}     & \multirow{2}{*}{DNN}                                                            &                                                                                                                   &                                                                                     \\ \cline{2-2} \cdashline{4-5}[1.2pt/1pt]
\multicolumn{1}{|c|}{}                                                                                      & \cite{alrabeiah2020deep}       &                                                                                 & \begin{tabular}[c]{@{}l@{}}$\bullet$  Geometry map\\ $\bullet$ CSI\end{tabular}                           & $\bullet$ Beamforming vector                                                                  \\ \hline
\multirow{3}{*}{\begin{tabular}[c]{@{}c@{}}Frequency\\ domain\\ prediction\end{tabular}}                    & \cite{wang2019ul}              & CNN+LSTM                                                                        & \multirow{3}{*}{$\bullet$ Uplink CSI}                                                                                  & \multirow{3}{*}{$\bullet$ Downlink CSI}                                                  \\ \cline{2-3}
                                                                                                            & \cite{arnold2019enabling}      & DNN                                                                             &                                                                                                                   &                                                                                     \\ \cline{2-3}
                                                                                                            & \cite{yang2020deep2}           & DT-meta-learning&                                                                                                                   &                                                                                     \\ \hline
\end{tabular}
\leftline{* Material factors, geometry map, and propagation factors may refer to different particular parameters in each paper, the }
\leftline{detail of each factors can be found in each reference.}
\leftline{* DNN here indicates a generalized deep learning network without using a specific structure, e.g., CNN.}
\end{table*}

\section{Important Issues and Challenges}

\subsection{Training Data Collection for AI-based Applications}
The training database is a critical point for designing/developing any ML method, i.e., the volume or the type of training data significantly affects the performance of AI-based applications regarding what type of ML method or what input features to use. The existing ML-based applications for wireless communications mostly use the training data collected from channel measurement to achieve the best performance in practical deployment. Nevertheless, the channel measurement is usually time-consuming and requires a high capital cost, and thus cannot be extensively conducted for massive scenarios. In this case, using synthetic data is an alternative way to build a training database.

There are mainly three ways to generate the synthetic data: 1) using the existing conventional models (such as standard models) or ML-based channel model; 2) using the simulator like RT; 3) using data reproduction method like GAN. The former two solutions can provide relatively accurate synthetic channel data in various scenarios, as a supplement to training data. In this case, the accuracy/performance of the trained ML method will highly depend on the accuracy of the models. In the other words, the performance of the ML-based applications is limited by the accuracy of the synthetic channels. Despite the limitations, the synthetic data is still a good compensation to some extension scenarios that conducting measurements are not feasible, especially for some extreme environment. On the other hand, the GAN-based simulator only reproduces the data that have the same inner pattern as training data (measurement data), and cannot provide new information but enhances the data pattern of the specific condition (the condition of the measurement data collection scenario). Hence, the synthetic generated by the GAN-based simulator cannot be considered as data of extension scenarios. In this sense, the synthetic data generated by the GAN-based simulator can be used as an enhancement if the volume of training data is not sufficient. Overall, both the measured data and the synthetic data should be tested as the training data for ML methods. Two major challenges are the limited volume of the measured data volume and the relatively low matching rate of the synthetic data. This leads to the crucial questions: i) how to improve the training efficiency by using a limited volume of the training data, and ii) how to improve the synthetic data accuracy compared to the measured data. These two challenges deserve further investigation in the future.

\subsection{AI-Based Channel Characterization for Positioning}
Future IoT networks and applications strongly rely on device positioning, which is still difficult for moving devices (e.g., smartphones or vehicles) in a complex environment with the continuous influence of human bodies and objects where sensors are mounted. But with the development of multi-antenna techniques and the computing power of the devices, more and more channel information can be obtained during communications. The contribution of the channel information to the localization/positioning has been extensively investigated \cite{burghal2020comprehensive}. According to the existing studies, the propagation environment impacts on the localization/positioning accuracy significantly. Therefore, the LoS/NLoS and even the physical scenario identification will greatly contribute to the localization/positioning performance. As reviewed in Section \ref{scenarioID}, many researchers studied the channel scenario identification based on static channel characteristics, but only a few research exploited the time-varying characteristic. As the matter of fact, the time-varying channel characteristics are sensitive to the changes of the propagation environment, in other words, the time-varying channel characteristics can significantly improve the efficiency and accuracy of the propagation scenario identification. However, there are still many critical problems that are not well addressed yet: how to exploit the time-varying channel characteristics to improve the positioning accuracy?  How to combine the scenario identification and the positioning? Which characteristics are more contribute to the scenario identification and positioning? All these questions still require further investigation.

\subsection{AI-Based Channel Prediction for Complicated/Combined Scenario}
As introduced in Section III, it is impractical to perform channel measurement campaigns everywhere. The question of the generalizability of channel models obtained from one environment to other environments needs to be investigated in more detail. To ensure the practicality of models for all similar scenarios, the accuracy is sacrificed in a way to achieve the trade-off between the generality and the specialness for a specific scenario.
Due to the flexibility of AI methods, the AI-based model can use more physical environmental factors as input to reveal the underlying connections between the channel properties and the physical propagation environment. This indicates that AI-based channel models can achieve better accuracy while maintaining the generality of their structure for different scenarios (since the changes of the physical environment are adopted as inputs), which makes it possible to predict the channel in the time/frequency/location domain based on the learning of channel history.  However, most of the existing research still separates the propagation environment into particular/representative scenarios, e.g., urban, suburban, or tunnel, before the modeling. The bridge between the physical environment and model parameters built by AI techniques brings the opportunity to build a complicated/combined scenario channel model that can flexibly support multiple scenarios with proper environmental factors and material factors, which can more accurately reconstruct/predict the experienced channels for moving terminals. For this purpose, the classification of scenarios, the representative and unified physical environmental factors for channel prediction will require further investigations. In addition, the transform between different scenarios usually leads to rapid changes of wireless channels, which is difficult to be modeled/predicted. Predicting the channels with rapid changes is still a challenging task for the prediction method. Besides, for prediction problems, the length of the time window of channel history for learning, the balance between the prediction accuracy and the prediction time length, and the selection of prediction method still require further study.

\subsection{AI-Based Channel Information Processing for Network Optimization}
As introduced in Section V, with the help of AI methods, highly accurate channel prediction becomes possible. Furthermore, with the development of the B5G and 6G communications, the high-speed movement scenario with ultra-low latency communication becomes one of the typical applications. This brings a big challenge for the data processing efficiency and the conventional network resource allocation methods may not meet the requirement of the new applications \cite{yang2020enabling, ye2019novel}. On the other hand, channel prediction is able to provide the CSI in advance, which gives more processing time for the network optimizations and ultimately leads to lower latency. In addition, as explained in Section \ref{scenarioID}, the channel scenario identification is not only an important precondition for high accuracy localization/positioning but also key information for network optimization. However, the channel modeling/prediction and the network optimization for scheduling/resource allocation are generally independent in the existing research. In this sense, the ML techniques can become the bridge of the two studies (since both have been tackled by ML).
Therefore, the merging of the channel scenario identification, channel prediction, and network optimization is one of the promising future areas of wireless communication system development.

\section{Conclusion}
AI techniques have become a necessary tool to develop the next generation communication network. In this paper, we provide a thorough overview of AI-enabled data processing for propagation channel studies, including the scenario identification and the channel modeling/prediction. This paper demonstrates the early results of the related works and illustrates the typical AI/ML-based solutions for each topic. Based on the state-of-art, the future challenges of AI/ML-based channel data processing techniques are given as well.

\bibliographystyle{ieeetr}
\bibliography{Overview_ref}

\begin{IEEEbiography}[{\includegraphics[width=1in,height=1.25in,clip,keepaspectratio]{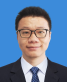}}]
{Chen Huang} (S'17-M'21) received the Ph.D. degrees from Beijing Jiaotong University, Beijing, China, in 2021. He is currently a research fellow in the Pervasive Communication Research Center, Purple Mountain Laboratories, Nanjing, 211111, China, and also a Post Doc in the National Mobile Communications Research Laboratory, School of Information Science and Engineering, Southeast University, Nanjing, 210096, China. His research interests are in machine-learning-based channel modeling.
\end{IEEEbiography}
\vspace{-1 cm}

\begin{IEEEbiography}[{\includegraphics[width=1in,height=1.25in,clip,keepaspectratio]{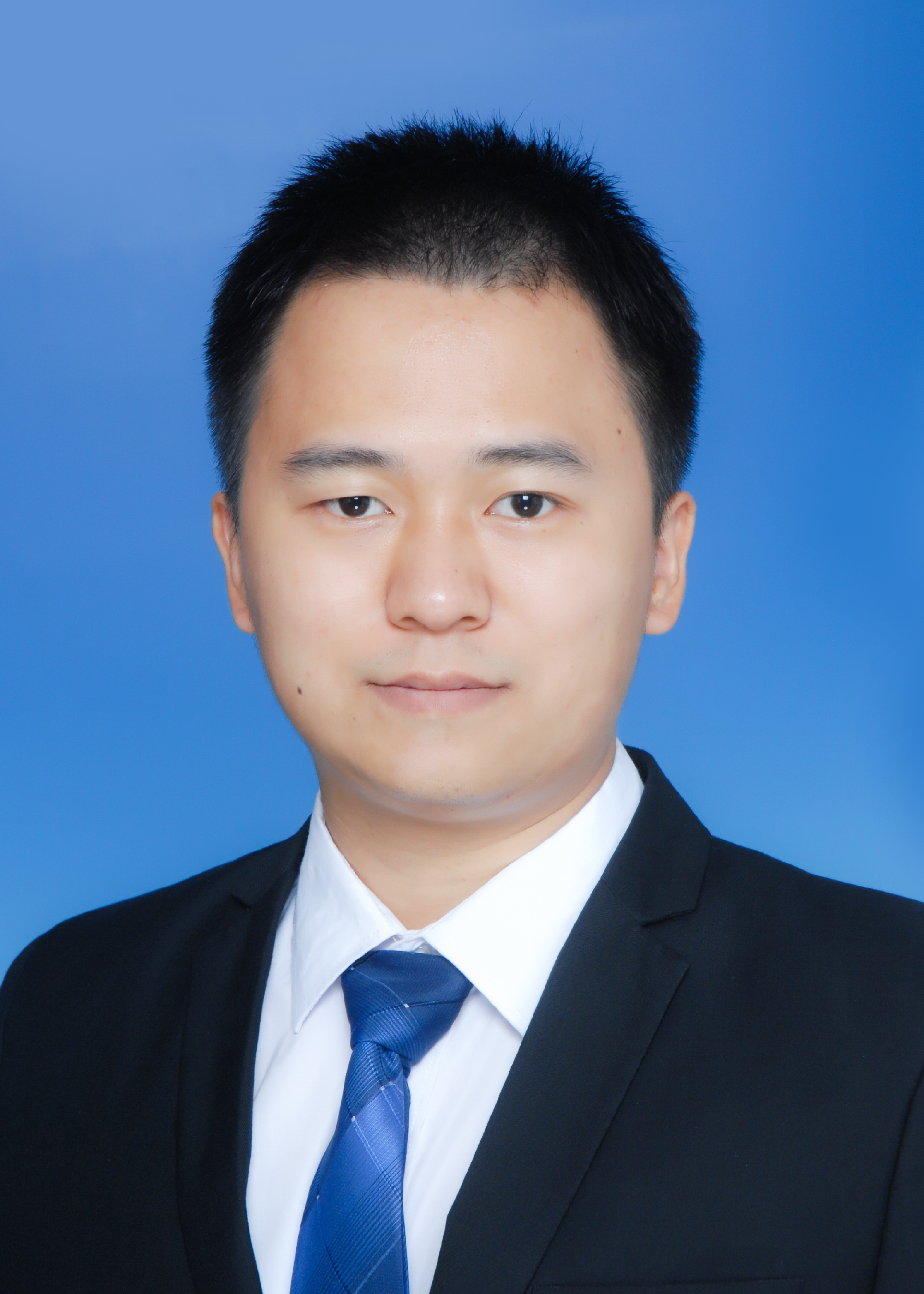}}]
{Ruisi He} (S'11-M'13-SM'17) received the B.E. and Ph.D. degrees from Beijing Jiaotong University, Beijing, China, in 2009 and 2015, respectively. Since 2015, Dr. He has been with the State Key Laboratory of Rail Traffic Control and Safety, BJTU, where he has been a Full Professor since 2018. Dr. He has been a Visiting Scholar in Georgia Institute of Technology, USA, University of Southern California, USA, and Universit\'e Catholique de Louvain, Belgium. His research interests include wireless propagation channels, railway and vehicular communications, 5G and 6G communications.
\end{IEEEbiography}
\vspace{-1 cm}

\begin{IEEEbiography}[{\includegraphics[width=1in,height=1.25in,clip,keepaspectratio]{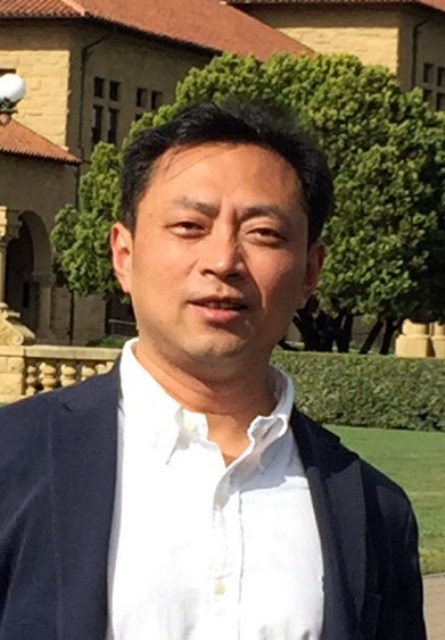}}]
{Bo Ai} (M'00-SM'10) is currently a Professor and an Advisor of Ph.D. candidates with Beijing Jiaotong University, Beijing, where he is also the Deputy Director of the State Key Laboratory of Rail Traffic Control and Safety. He is also currently with the Engineering College, Armed Police Force, Xian.  His interests include the research and applications of orthogonal frequency-division multiplexing techniques, high-power amplifier linearization techniques, radio propagation and channel modeling, global systems for mobile communications for railway systems, and long-term evolution for railway systems.
\end{IEEEbiography}
\vspace{-1 cm}

\begin{IEEEbiography}[{\includegraphics[width=1in,height=1.25in,clip,keepaspectratio]{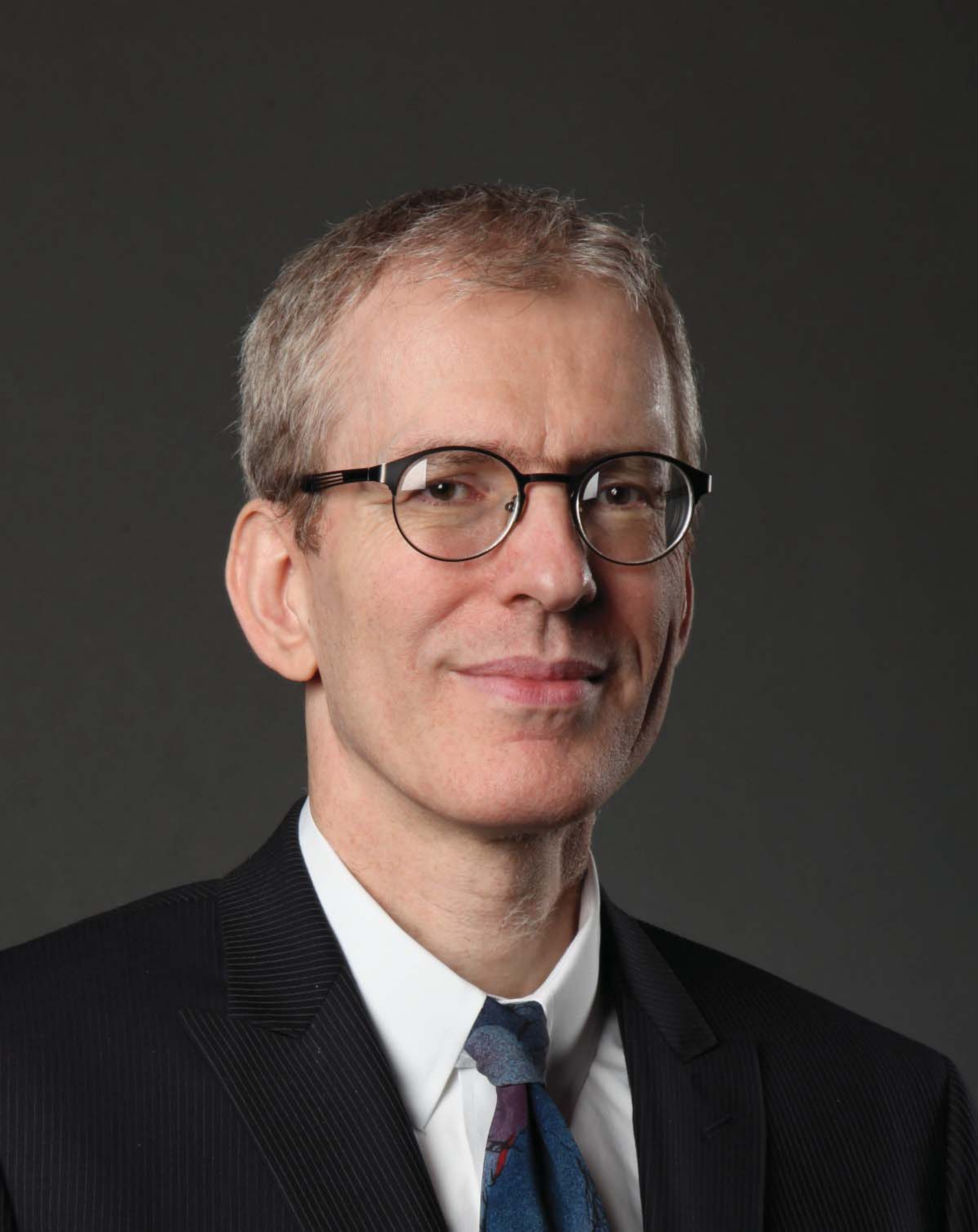}}]
{Andreas F. Molisch} (S'89-M'95-SM'00-F'05)  is the Solomon Golomb Andrew and Erna Viterbi Chair Professor at the University of Southern California. He was previously at TU Vienna, AT\&T (Bell) Labs, Lund University, and Mitsubishi Electric Research Labs. His research interests are in wireless communications, with emphasis on propagation channels, multiantenna systems, ultrawideband systems, and localization. He is a Fellow
of NAI, AAAS, and IET, a member of the Austrian Academy of Sciences, and a recipient of numerous awards.
\end{IEEEbiography}
\vspace{-1 cm}

\begin{IEEEbiography}[{\includegraphics[width=1in,height=1.25in,clip,keepaspectratio]{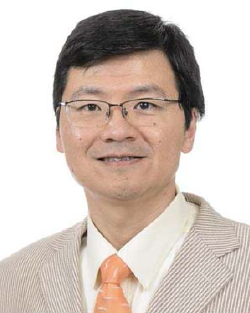}}]
{Buon Kiong Lau}(Senior Member, IEEE) received the B.E. degree (Hons.) in electrical engineering from the University of Western Australia, Perth, WA, Australia, in 1998, and the Ph.D. degree in electrical engineering from the Curtin University of Technology, Perth, in 2003. Since 2004, he has been with the Department of Electrical and Information Technology, Lund University, where he is currently a Professor. His primary research interests are in various aspects of multiple antenna systems, particularly the interplay between antennas, propagation channels, and signal processing.
\end{IEEEbiography}
\vspace{-1 cm}

\begin{IEEEbiography}[{\includegraphics[width=1in,height=1.25in,clip,keepaspectratio]{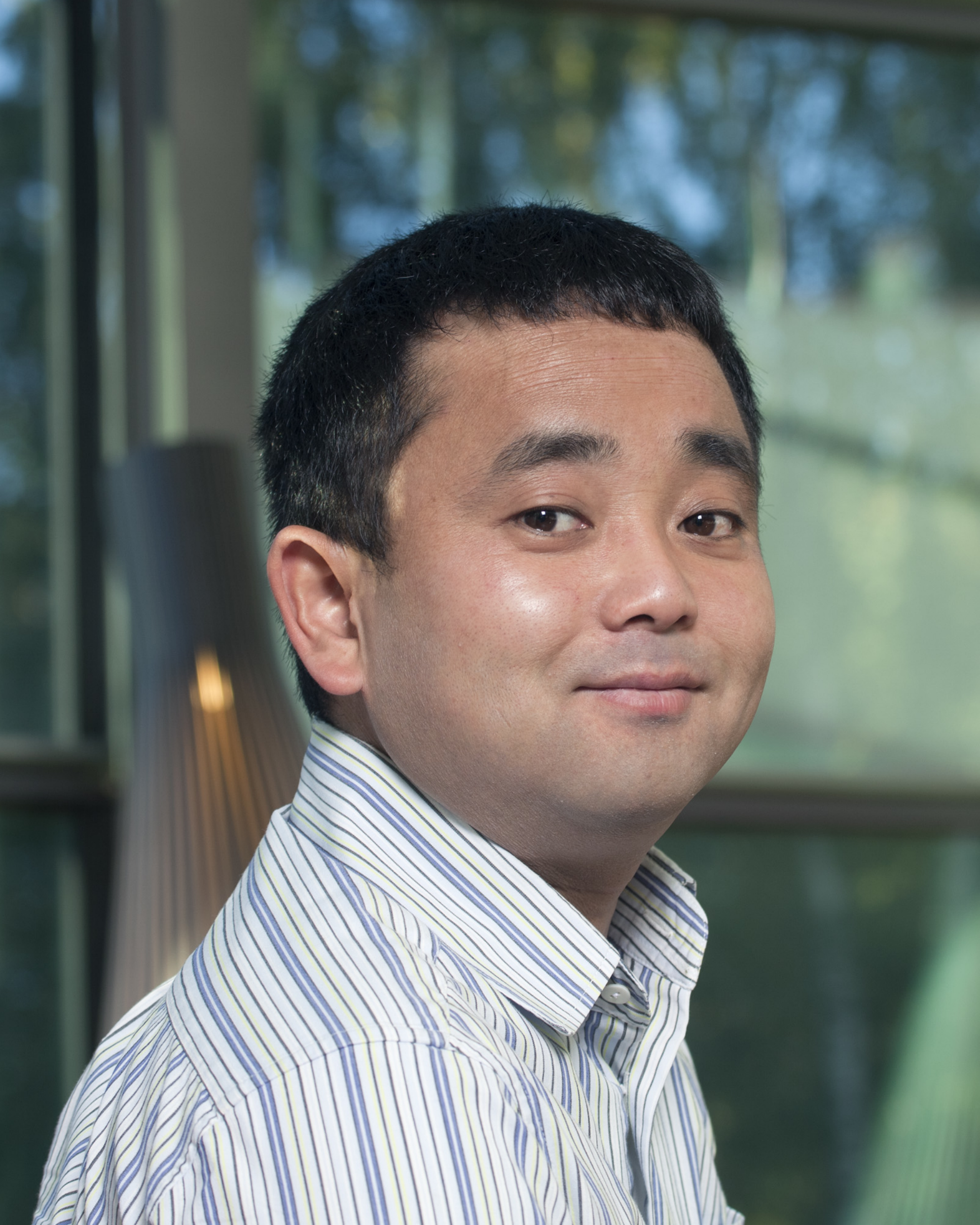}}]
{Katsuyuki Haneda} is an associate professor in the Aalto University, Finland. He has been an associate editor of the IEEE Transactions on Antennas and Propagation for 2012-2016, and of an editor of the IEEE Transactions on Wireless Communications for 2013-2018. His current research activity includes radio frequency instrumentation, measurements and modeling, millimeter-wave radios, in-band full-duplex radio technology and radio applications in medical and healthcare scenarios.
\end{IEEEbiography}
\vspace{-1 cm}

\begin{IEEEbiography}[{\includegraphics[width=1in,height=1.25in,clip,keepaspectratio]{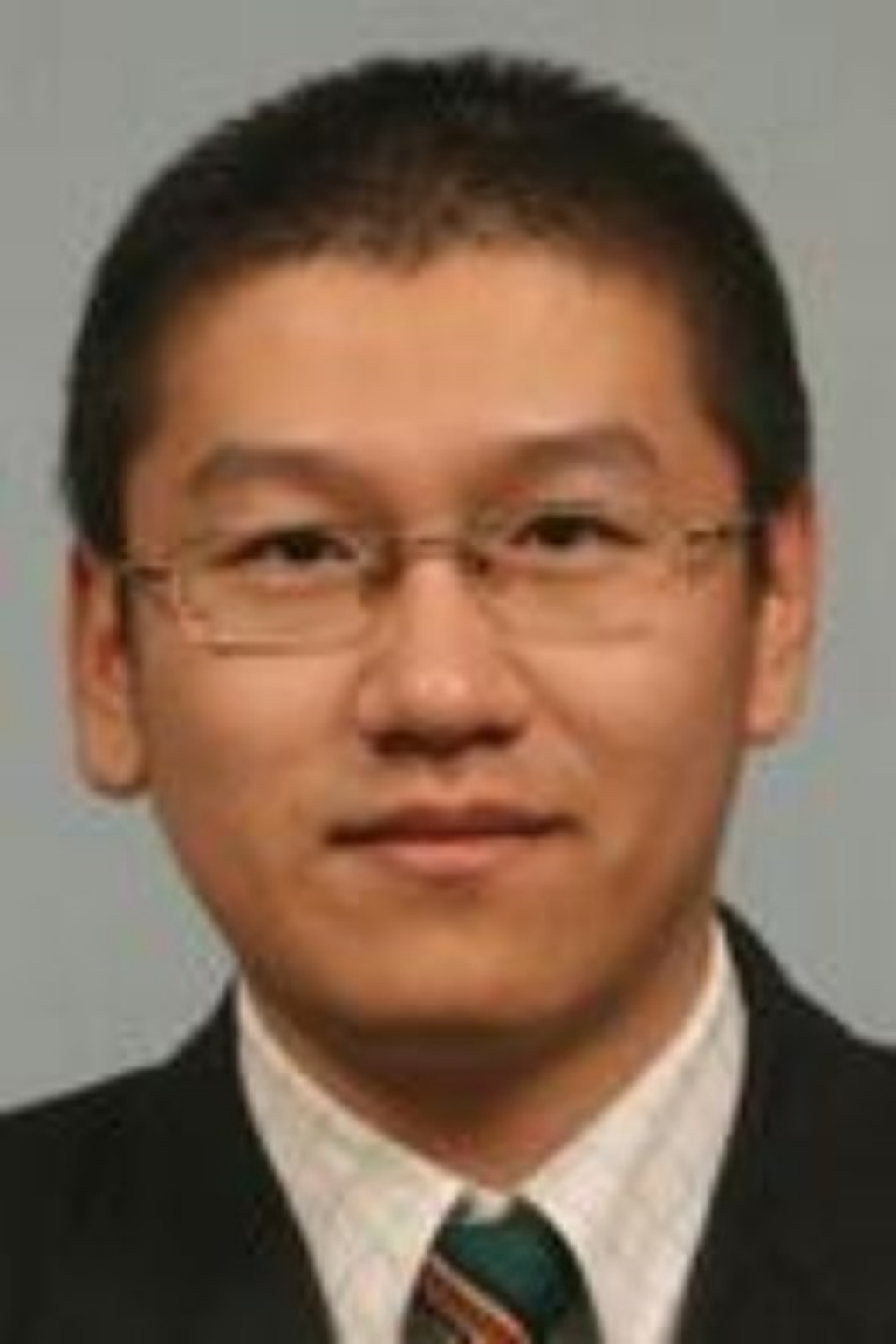}}]
	{Bo Liu}(M'15-SM'17) received the B.S. degree from Tsinghua University, Beijing, China, in 2008, and the Ph.D. degree from the University of Leuven (KU Leuven), Leuven, Belgium, in 2012. He is currently a Senior Lecturer (Associate Professor) with the University of Glasgow, Glasgow, U.K. He is also a Senior Honorary Fellow with the University of Birmingham, Birmingham, U.K.. His research interests lie in artificial intelligence-driven design methodologies of analog/RF integrated circuits, microwave devices, MEMS, evolutionary computation, and machine learning.
\end{IEEEbiography}
\vspace{-1 cm}

\begin{IEEEbiography}[{\includegraphics[width=1in,height=1.25in,clip,keepaspectratio]{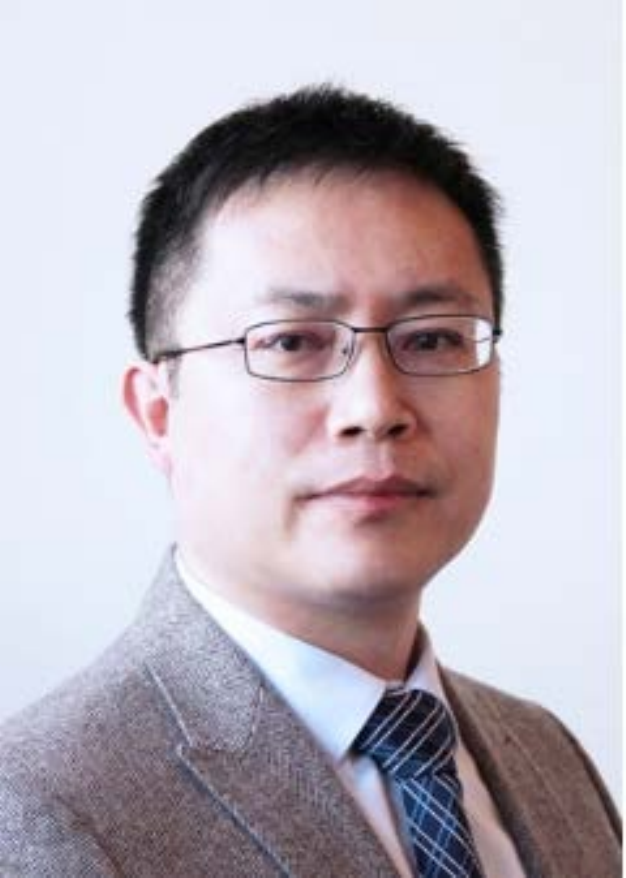}}]
	{Cheng-Xiang Wang}(S'01-M'05-SM'08-F'17) received the BSc and MEng degrees in Communication and Information Systems from Shandong University, China, in 1997 and 2000, respectively, and the PhD degree in Wireless Communications from Aalborg University, Denmark, in 2004. In 2018, he joined the National Mobile Communications Research Laboratory, Southeast University, China, as a Professor. He is also a parttime professor with the Purple Mountain Laboratories, Nanjing, China. His current research interests include wireless channel measurements and modeling, B5G wireless communication networks, and applying artificial intelligence to wireless networks.
 \end{IEEEbiography}
 \vspace{-1 cm}

\begin{IEEEbiography}[{\includegraphics[width=1in,height=1.25in,clip,keepaspectratio]{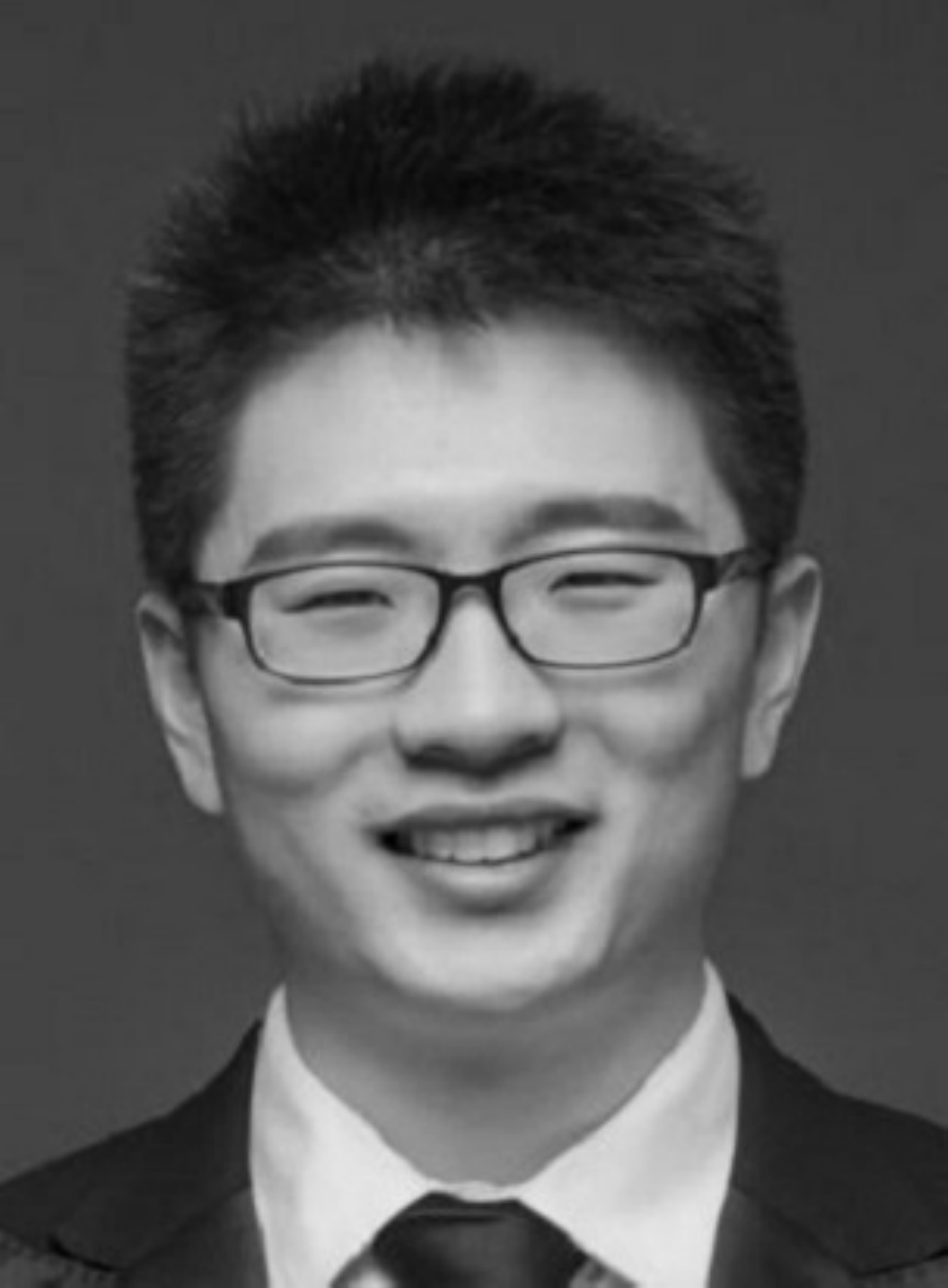}}]
{Mi Yang} (S'17-M'21) received the M.S. and Ph.D. degrees from Beijing Jiaotong University (BJTU), Beijing, China, in 2017 and 2021, respectively. He is currently a associate professor with the State Key Laboratory of Rail Traffic Control and Safety, Beijing Jiaotong University. His research interests are focused on wireless propagation channels, vehicle-to-everything (V2X) communications and 5G/B5G communications.
\end{IEEEbiography}
\vspace{-1 cm}

\begin{IEEEbiography}[{\includegraphics[width=1in,height=1.25in,clip,keepaspectratio]{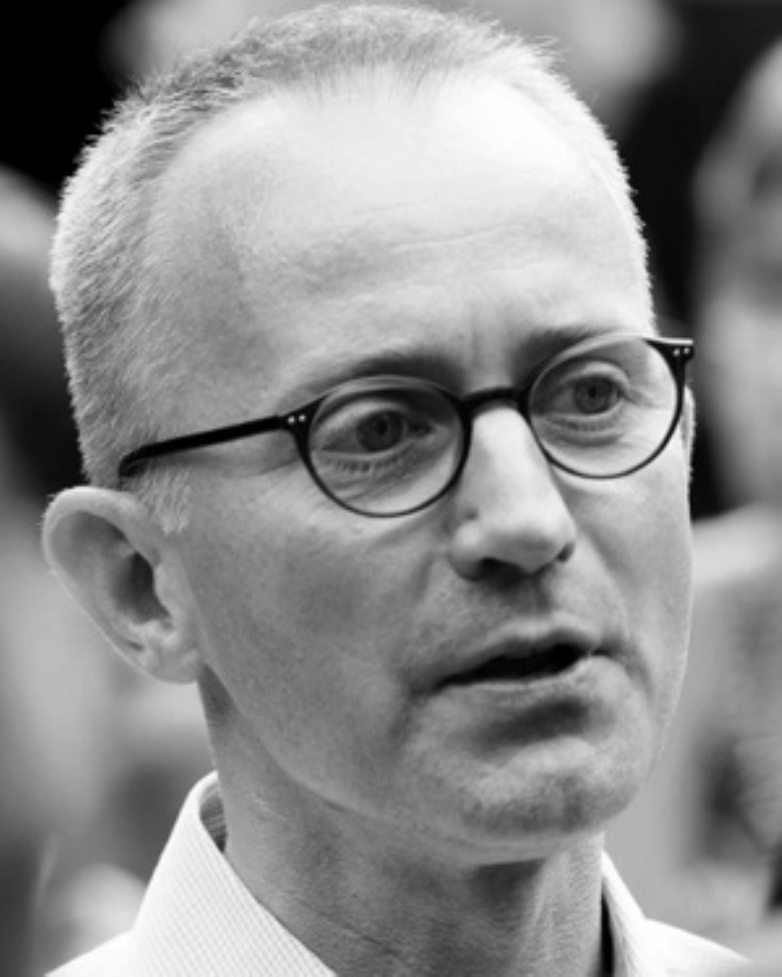}}]
	{Claude Oestges}(Fellow, IEEE) received the M.Sc and Ph.D. degrees in Electrical Engineering from the  Universit\'e Catholique de Louvain (UCLouvain), \'Louvain-la-Neuve, Belgium, in 1996 and 2000, respectively. In January 2001, he joined the Smart Antennas Research Group (Information Systems Laboratory), Stanford University, Stanford, CA, USA, as a Postdoctoral Scholar. From January 2002 to September 2005, he was associated with the Microwave Laboratory, UCLouvain, as a Postdoctoral Fellow of the Belgian Fonds de la Recherche Scientifique (FRS-FNRS). He is currently a Full Professor with the Electrical Engineering Department, Institute for Information and Communication Technologies, Electronics and Applied Mathematics (ICTEAM), UCLouvain.
\end{IEEEbiography}
\vspace{-1 cm}

\begin{IEEEbiography}[{\includegraphics[width=1in,height=1.25in,clip,keepaspectratio]{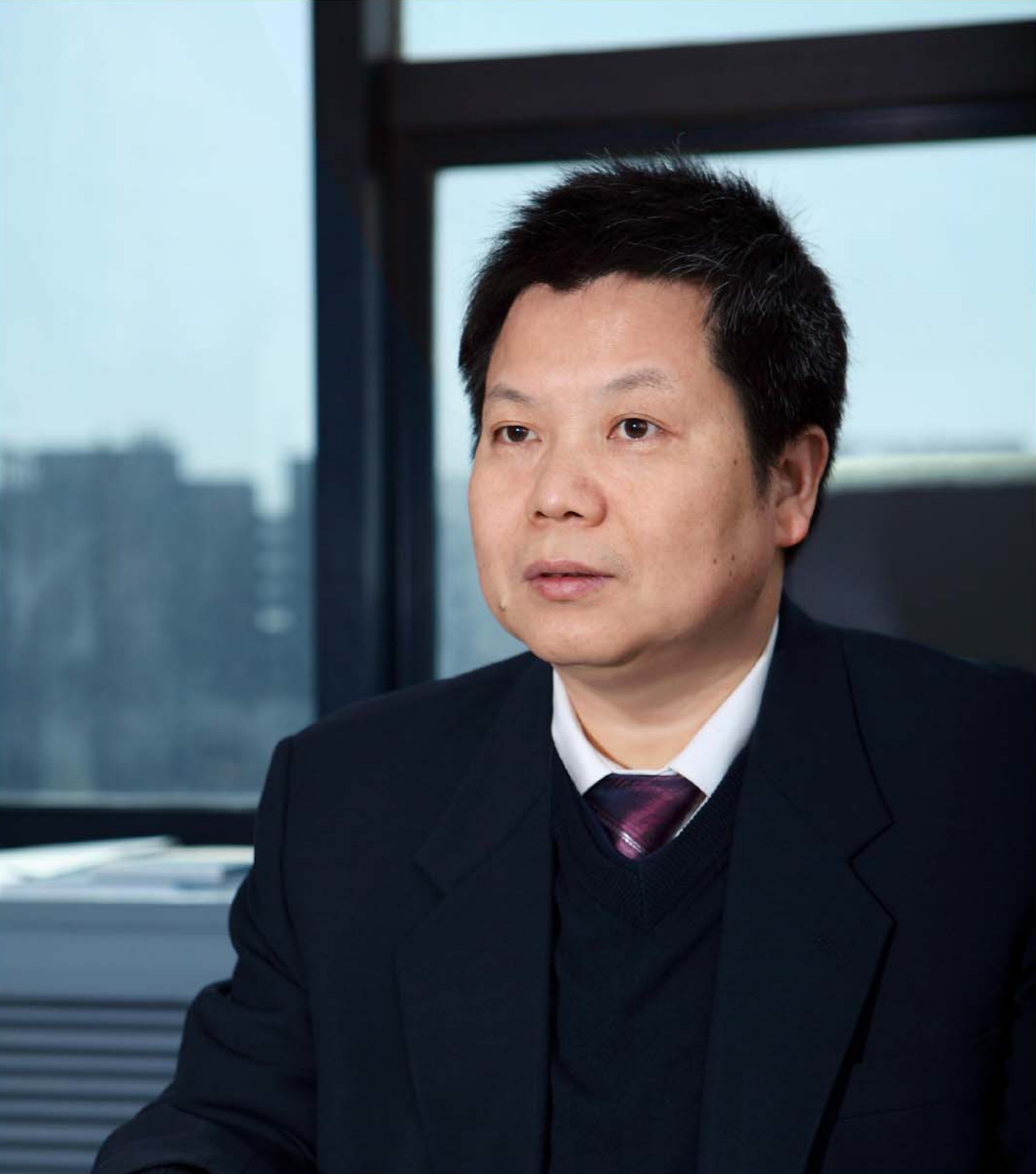}}]
	{Zhangdui Zhong} (SM'16) received the B.E. and M.S. degrees from Beijing Jiaotong University, Beijing, China, in 1983 and 1988, respectively. He is a Professor and Advisor of Ph.D. candidates with Beijing Jiaotong University, Beijing, China. He is currently a Director of the School of Computer and Information Technology and a Chief Scientist of State Key Laboratory of Rail Traffic Control and Safety, Beijing Jiaotong University. He is an Executive Council Member of Radio Association of China, Beijing, and a Deputy Director of Radio Association, Beijing. His interests include wireless communications for railways, control theory and techniques for railways, and GSM-R systems.
\end{IEEEbiography}

%
%
%

%
%
%
%
%




\end{document}